\documentclass[12pt]{article}
\usepackage{amsmath}
\usepackage{mathtools} 
\usepackage{amsfonts}
\usepackage{amssymb}
\usepackage{geometry}
\usepackage{graphicx}
\usepackage{fancyhdr}
\usepackage{titlesec}
\usepackage{tikz-cd}
\usepackage{tikz}
\usetikzlibrary{cd, arrows.meta, positioning}

\tikzset{
  alg/.style = {draw, rectangle, minimum width=3cm, align=center, minimum height=1.2cm},
  opt/.style = {draw, thick, rectangle, minimum width=6cm, align=center, minimum height=1.4cm},
  heur/.style = {draw, rectangle, minimum width=3cm, align=center, minimum height=1.0cm},
  arrow/.style = {thick, ->, >=Stealth, shorten >=5pt, shorten <=5pt},
}

\usepackage{url}

\begin{document}

\begin{titlepage}
    \centering
    \vspace*{0cm} 
    {\Large\textbf{Introducing Image-Space Preconditioning in the Variational Formulation of MRI Reconstructions}\par}
    \vskip0.4cm
    { \textbf{\small{Bastien Milani}}${}^{1}$, \textbf{\small{Jean-Baptist Ledoux}}${}^{2, 3}$, \textbf{\small{Berk Can Açikgöz}}${}^{4, 5, 6}$, \textbf{\small{Xavier Richard}}${}^{1}$\par}
    \vskip0.4cm
    \small${}^{1}$\small{HES-SO Valais Wallis, Sion, Switzerland}\\
    \small${}^{2}$\small{CIBM Center for Biomedical Imaging, Lausanne, Switzerland} \\
    \small${}^{3}$\small{Department of Diagnostic and Interventional Radiology, Lausanne, Switzerland} \\ 
    \small${}^{4}$\small{Graduate School for Cellular and Biomedical Sciences, University of Bern, Bern, Switzerland} \\
    \small${}^{5}$\small{Department of Diagnostic, Interventional and Pediatric Radiology (DIPR), Swiss Institute for Translational and Entrepreneurial Medicine, Bern, Switzerland} \\
    \small${}^{6}$\small{Translational Imaging Center (TIC), University of Bern, Bern, Switzerland} \\
    \vskip0.3cm
    \small{\textit{Correspondence: Bastien Milani, Email: bastien.milani@hes-so.ch}}
    \vskip0.4cm

    \begin{abstract}
    The aim of the present article is to enrich the comprehension of iterative magnetic resonance imaging (MRI) reconstructions, including compressed sensing (CS) and iterative deep learning (DL) reconstructions, by describing them in the general framework of finite-dimensional inner-product spaces. In particular, we show that image-space preconditioning (ISP) and data-space preconditioning (DSP) can be formulated as non-conventional inner-products. The main gain of our reformulation is an embedding of ISP in the variational formulation of the MRI reconstruction problem (in an algorithm-independent way) which allows in principle to naturally and systematically propagate ISP in all iterative reconstructions, including many iterative DL and CS reconstructions where preconditioning is lacking. The way in which we apply linear algebraic tools to MRI reconstructions as presented in this article is a novelty.
    
    A secondary aim of our article is to offer a certain didactic material to scientists who are new in the field of MRI reconstruction. Since we explore here some mathematical concepts of reconstruction, we take that opportunity to recall some principles that may be understood for experts, but which may be hard to find in the literature for beginners. In fact, the description of many mathematical tools of MRI reconstruction is fragmented in the literature or sometimes missing because considered as a general knowledge. Further, some of those concepts can be found in mathematic manuals, but not in a form that is oriented toward MRI. For example, we think of the conjugate gradient descent, the notion of derivative with respect to non-conventional inner-products, or simply the notion of adjoint. The authors believe therefore that it is beneficial for their field of research to dedicate some space to such a didactic material. 
    \end{abstract}

    
\end{titlepage}

\section{Introduction}

The variational formulation of the reconstruction problems plays a crucial role in modern MRI reconstruction methodologies by providing a unified framework for translating algorithmic approaches into optimization problems. The first part of this introduction will focus on the historical development of this transition, highlighting its importance. A critical aspect of this evolution has been the conceptual shift between reconstruction algorithms and optimization-based formulations, which has facilitated the creation of new MRI reconstruction techniques. This transition was initiated by Pruessmann et al (2001) \cite{iterSense}, who formulated the reconstruction problem as a normal equation without preconditioning or a (almost) normal equation with preconditioning. Since a normal equation is mathematically equivalent to its corresponding least squares (LS) problem - a convex optimization problem - this formulation implicitly established a connection between reconstruction algorithms and LS optimization problems. 

The formalization of this link appeared shortly thereafter in the work of Sutton, Noll, and Fessler in 2003 \cite{suttonLS}, who explicitly wrote the MRI reconstruction problem in terms of a least-squares minimization, albeit without referring to its equivalence with the normal equations derived in \cite{iterSense}. Whether this formulation emerged independently or not, the equivalence between the two approaches remains a mathematical fact. We will therefore regard the least-squares problem proposed in \cite{suttonLS} as the variational formulation of the normal equations introduced in \cite{iterSense}. This recognition—whether or not it was made explicitly by the authors—proved to be a crucial step, as it opened the door to a natural extension of the framework: the formulation of the regularized least-squares reconstruction problem, also presented in \cite{suttonLS}.
Recasting the reconstruction problem in this abstract variational framework not only unified algorithmic and optimization-based approaches, but also laid the theoretical foundation for the iterative algorithms that forms one major branch contemporary MRI reconstructions.

In the second part of this introduction, we highlight a critical omission in the variational formulation: image-space preconditioning (ISP). This essential component is notably absent from the variational formulations employed in reconstruction problems. The incorporation of ISP into the optimization problem has significant implications for both compressed sensing (CS) reconstructions \cite{reviewCS1, reviewCS2} and iterative deep learning (DL) reconstructions \cite{vSharp, yiasemis2022, surveyDL}, as many of the modern algorithms are based on classical methods that do not inherently incorporate preconditioning. We propose two approaches to integrate ISP into the optimization framework. The second method is particularly elegant, utilizing the flexibility in selecting an inner-product within image-space to encapsulate the effects of ISP directly within this product. 

In the third part of this introduction, we present our main contribution: the integration of image-space Preconditioning (ISP) into the variational framework of MRI reconstruction. We revisit the historical trajectory by re-entering through the “iterative SENSE” approach—precisely where ISP was left behind when the field shifted toward variational methods. This allows us to systematically reintroduce ISP into algorithms derived from such formulations. Concurrently, we also reformulate data-space preconditioning (DSP), also called k-space preconditioning \cite{ong2020}, by incorporating it directly into the inner product of data-space. This reveals a natural symmetry between image-space and data-space, enabling a unified interpretation of iterative reconstruction methods within the framework of inner-product spaces.

This perspective reinforces the paradigm of transitioning between algorithmic and variational viewpoints to generate new reconstruction strategies. Notably, we show that the image-intensity correction introduced in \cite{iterSense} arises as a special case of ISP in our framework. By embedding ISP into the variational formulation, we enable its propagation to a wide range of methods, including compressed sensing (CS) reconstructions where it is often absent \cite{noPrecond1, noPrecond2, noPrecond3, noPrecond4}, and iterative deep-learning approaches, which to our knowledge have never incorporated ISP.

\subsection{Some elements of the historical pathway \newline to regularized least-square reconstructions}

In the following, we define $X \simeq\mathbb{C}^{nVox}$ as the image-space. This vector space encompasses all possible MRI images with a total number of voxels equal to $nVox$. Similarly, we denote $Y\simeq\mathbb{C}^{nSamp}$ as the data-space, representing all possible measured data for a given number of samples $nSamp$.
We consider an iterative reconstruction algorithm as a specific implementation that generates a reconstructed image $x^\# \in X$ based on some input measured data $y_0 \in Y$ and an initial image guess $x_0$. Generally, the reconstructed image depends on the choice of the initial guess $x_0$, but the set of images that can be produced by a given algorithm for a specific dataset is a subset of $X$.
If this subset coincides with the set of minimizer of a particular function $\Gamma(\cdot)$ for some given data $y_0$, we can associate the algorithm with an optimization problem. This association implies that the iterative reconstruction process aims to find solutions within the framework defined by the objective function $\Gamma(\cdot)$.

\begin{equation*}
\text{Find } x^\# \in \underset{x \in X}{\operatorname{argmin}} \  \Gamma(x)
\end{equation*}

In this work, we emphasize the intrinsic link between an algorithm and its corresponding optimization problem. Once a reconstruction method is formulated as an optimization problem, it becomes possible to explore alternative algorithms derived from this formulation. These new algorithms may offer improved performance or efficiency compared to the original approach.

\[
\begin{tikzcd}[row sep=2cm, column sep=2.5cm]
  \node[alg] (A1) {New Algorithm 1};
  \node[opt, right=of A1] (OPT) {$x^\# \in \displaystyle\arg\min_{x \in X} \Gamma(x)$};
  \node[alg, right=of OPT] (A2) {New Algorithm 2};
  \node[alg, below=of OPT] (A0) {Native Algorithm};

  \draw[arrow] (A1) -- (OPT);
  \draw[arrow] (A2) -- (OPT);
  \draw[arrow] (A0) -- (OPT);
  \draw[arrow] (OPT) -- (A1);
  \draw[arrow] (OPT) -- (A2);
  \draw[arrow] (OPT) -- (A0);
\end{tikzcd}
\]

Furthermore, modifying the objective function $\Gamma(\cdot)$—for example by adding or adjusting a regularization term—and then deriving algorithms to solve the new problem can lead to novel algorithmic strategies. These would be difficult to obtain directly from the original method but become accessible through the variational perspective.

\[
\begin{tikzcd}[row sep=2cm, column sep=2.8cm]
  \node[opt] (A) {$x^\# \in \displaystyle\arg\min_{x \in X} \Gamma(x)$};
  \node[opt, right=of A] (B) {$x^\# \in \displaystyle\arg\min_{x \in X} \Gamma(x) + \lambda R(x)$};

  \node[alg, below=of A] (C) {Native Algorithm};
  \node[alg, below=of B] (D) {New Algorithms};

  \node[heur, below=of D] (E) {Heuristic Transition};
  \node[alg, below=of E] (F) {\small Trained Algorithm};

  \draw[arrow] (A) -- (C);
  \draw[arrow] (B) -- (D);
  \draw[arrow] (D) -- (E);
  \draw[arrow] (E) -- (F);

  \draw[arrow] (A) -- node[above]{ \text{Easy}} (B);
  \draw[arrow] (C) -- node[above]{ \text{Difficult}} (D);

  \draw[arrow] (C) -- (A);
  \draw[arrow] (D) -- (B);
\end{tikzcd}
\]

These new algorithms may then undergo heuristic modifications, such that they no longer directly correspond to an optimization problem but can instead be trained on data to improve reconstruction performance. We refer to this process as a \textit{heuristic transition}. 

This pattern, transitioning between algorithmic and variational paradigms, introducing modifications along the way, and returning with enriched formulations, has shown to be very efficient in the context of MRI reconstruction. That path was initiated by the introduction of iterative SENSE \cite{iterSense}, which formulated the reconstruction problem initially (Eq. 13 in \cite{iterSense}) as a normal equation:

\begin{equation*}
E^* E x^\# = E^* y_0 
\end{equation*}
and later (Eq. 24 in \cite{iterSense}) as:
\begin{equation*}
\left( I E^* D E I \right) \left( I^{-1} x^\# \right) = I E^* D y_0 
\end{equation*}
where $I$ denotes the image-intensity correction matrix, $D$ the (k-space) density compensation matrix, $E$ is the matrix of the MRI forward model, and $E^*$ is its complex-conjugate transpose matrix. While the first equation corresponds to a standard normal equation, we will show below that the second can likewise be reformulated as such. Accordingly, we consider that the reconstruction problem in \cite{iterSense} was indeed cast in the form of a normal equation.

Since any normal equation is fully equivalent to its associated least-squares (LS) formulation, the iterative-SENSE framework can be viewed as an early instance of a variational approach. This variational perspective was subsequently made explicit by Sutton, Noll, and Fessler \cite{suttonLS}. Although the LS problem was not written in closed form in their article, it was clearly articulated in the text. We restate it here as:
\begin{equation*}
\text{Find }x^\# \in \underset{x \in X}{\operatorname{argmin}} \frac{1}{2} \| E x - y_0 \|_2^2
\end{equation*}
To the best of our knowledge, the article \cite{suttonLS} marked the first explicit variational formulation of the MRI reconstruction problem. This milestone, though significant, received limited attention: even today, many works refer to the LS formulation of the MRI reconstruction problem without acknowledging its origin.

By casting the MRI reconstruction problem as an LS problem, the authors of \cite{suttonLS} enabled the addition of a regularization term, yielding the regularized least-squares (RLS) formulation:
\begin{equation*}
\text{Find }x^\# \in \underset{x \in X}{\operatorname{argmin}} \frac{1}{2} \| E x - y_0 \|_2^2 + \lambda R(x)
\end{equation*}
where $\lambda > 0$ is a regularization parameter, $R(\cdot)$ a real-valued functional, and $\lambda R(x)$ the regularization term. The $l_2$ norm $\lVert \cdot \rVert_2$ denotes the conventional Euclidean norm. Regularization serves to reduce noise, suppress undersampling artifacts, and eliminate undesired solutions in ill-posed settings. The same article also proposed algorithms to solve specific $l_2$-regularized cases.

In the years that followed, numerous variants of the RLS problem and associated algorithms were explored \cite{REGLS1, REGLS2, REGLS3}, with a notable subfamily being compressed sensing (CS) reconstructions \cite{reviewCS1, reviewCS2}. Introduced in MRI by Lustig \cite{lustigCS} and grounded in the CS theory of Donoho \cite{donohoCS} and Candès et al. \cite{candesCS}, these reconstructions are based on $l_1$-regularization.

More recently, deep-learning (DL) techniques have demonstrated strong performance in MRI reconstruction \cite{vSharp, yiasemis2022, surveyDL}. For iterative schemes, these methods typically consist in modifying non-trained algorithms by replacing update steps with learned operators. To the author’s knowledge, all iterative reconstructions operating in image space—including CS and iterative DL methods—ultimately originate from the RLS formulation.

This trajectory illustrates how an abstract variational formulation can lead to tangible and powerful developments. In view of its central role, we refer to the RLS formulation hereafter as the reconstruction problem.

\[
\begin{tikzcd}[row sep=2.2cm, column sep=3.2cm]
  \node[opt] (A) {$x^\# \in \displaystyle\arg\min_{x \in X} \frac{1}{2} \| E x - y_0 \|_2^2$};
  \node[opt, right=of A] (B) {$x^\# \in \displaystyle\arg\min_{x \in X} \frac{1}{2} \| E x - y_0 \|_2^2 + \lambda R(x)$};

  \node[alg, below=of A] (C) {Iterative-SENSE Algorithm};
  \node[alg, below=of B] (D) {\small Compressed-Sensing Algorithms\\+ others};

  \node[alg, below=of C] (E) {\scriptsize SENSE, GRAPPA, SMASH,\\ \scriptsize VD-AUTO-SMASH, PILS, ...};
  \node[heur, below=of D] (F) {Heuristic Transition};

  \node[alg, below=of F] (G) {\small Iterative Deep-Learning Reconstruction};

  \draw[arrow] (A) -- (C);
  \draw[arrow] (B) -- (D);
  \draw[arrow] (C) -- (E);
  \draw[arrow] (D) -- (F);
  \draw[arrow] (F) -- (G);

  \draw[arrow] (A) -- node[above]{\text{\small Easy}} (B);
  \draw[arrow] (C) -- node[above]{\text{\small Difficult}} (D);

  \draw[arrow] (C) -- (A);
  \draw[arrow] (D) -- (B);
\end{tikzcd}
\]

Iterative-SENSE thus served as a gateway to the abstract variational formulation by implicitly assuming that the reconstruction problem could be expressed as a normal equation. Importantly, it also acted as a unifying framework for all single-frame, non-iterative parallel imaging methods of that time. Indeed, prior to iterative-SENSE, techniques such as SENSE \cite{sense}, GRAPPA \cite{grappa}, SMASH \cite{smash}, VD-AUTO-SMASH \cite{vdautosmash}, PILS \cite{pils}, and others, were all non-iterative approaches for static imaging that effectively solved the same problem as iterative-SENSE, but only for specific sampling schemes.

Iterative-SENSE generalized these methods to arbitrary Cartesian and non-Cartesian sampling patterns. Consequently, all these reconstruction strategies—including iterative-SENSE—can be associated with the same underlying least-squares formulation. Symbolically, one could say that all earlier parallel imaging methods converged toward the same variational formulation, with iterative-SENSE opening the door to this unified perspective.

This historical role forms the cornerstone of the present study: the iterative-SENSE algorithm will serve as a reference baseline for introducing a key, previously overlooked component within the variational framework—namely, image-space preconditioning (ISP).

\subsection{The Missing Piece}
The notion of image intensity correction, introduced in \cite{iterSense}, constitutes the earliest example of what we call here "image-space preconditioning" (ISP), also called simply “preconditioning” in \cite{ong2020}. Image intensity correction is introduced in \cite{iterSense} as a diagonal matrix $I$ inserted in the linear reconstruction equation and propagates from there into the reconstruction algorithm. The same publication introduced the notion of k-space density compensation which is the first example of data-space preconditioning (DSP), also call k-space preconditioning. As for image intensity correction, k-space density compensation is introduced in the reconstruction equation as a diagonal matrix, written $D$, and also propagates from there in the reconstruction algorithm. But since no variational formulation of the reconstruction problem is expressed in \cite{iterSense}, neither ISP nor DSP is expressed in a variational problem. 

DSP was then introduced in the variational setting quite early after its first publication. This can be done by replacing $E$ with $\sqrt{D} \ E$ and by replacing $y_0$ with $\sqrt{D}\ y_0$, where $\sqrt{D}$ satisfies $\sqrt{D}^2 = D$. These two substitutions are equivalent to a change of coordinates in the data-space and is typically introduced to accelerate convergence. Alternatively, DSP can be embedded in the variational problem of the reconstruction problem by inserting the matrix $\sqrt{D}$ as follows: 
\begin{equation*}
\text{Find }x^\# \in \underset{x \in X}{\operatorname{argmin}} \frac{1}{2} \| \sqrt{D} \ E x - \sqrt{D} \ y_0 \|_2^2 + \lambda R(x)
\end{equation*}

Other kind of preconditioning (other than DSP) have later been implemented in reconstructions algorithms \cite{suttonLS, ong2020, ramaniPrecond, wellerPrecond, koolstraPrecond}. Some of those preconditioning techniques can be qualified as ISP while other are neither ISP nor DSP. But independently of their nature, and to the exception of DSP, none of those preconditioning techniques have been embedded in the variational problem of MRI reconstruction. DSP is the only one. The other implementations of preconditioning (including ISP) were introduced as a heuristic modification of the algorithm to improve convergence speed and remained tightly coupled to specific algorithms, without being integrated into the variational problem itself. It follows in particular for ISP, the focus of the present article, that it has never been embedded in the variational problem and appears therefore as a missing piece in the variational formulation, which breaks the symmetry between data-space and image-space. 

We want here to restore that symmetry and introduce ISP in the variational problem, just like DSP. This can be performed quite easily by a change of coordinates similarly to DSP: we define the substitute variable $\tilde{x} = I^{-1} x$ living in the space $\tilde{X} \simeq \mathbb{C}^{nVox}$ and we note that
\begin{equation*}
\sqrt{D} \ E x  = \sqrt{D} \ E \ I \ \left(I^{-1} x \right) =  \sqrt{D} \ E \ I \ \tilde{x}
\end{equation*}
The variational formulation of the reconstruction problem, with ISP embedded, can then be written as 
\begin{equation*}
\text{Find }x^\# \in I \cdot \underset{\tilde{x} \in \tilde{X}}{\operatorname{argmin}} \frac{1}{2} \| \sqrt{D} \ E \ I \tilde{x} - \sqrt{D} \ y_0 \|_2^2 + \lambda R(I \tilde{x})
\end{equation*}
At that point, the reader may probably raise the two following questions: if embedding ISP in the variational problem is so easy, 
\begin{itemize}
    \item a) why did ISP not appear in the variational formulation simultaneously to DSP ? 
    \item b) what is the sense of writing a long article about it ? 
\end{itemize}
The author ignores the answer to the first question and can only speculate about it (see discussion section). But it may be that working with a substitute variable instead of the image itself may have been counter intuitive for MRI reconstruction scientists, and it may be that they just decided avoid it. The answer to the second question is the reason of the existence of the present article : much more than introducing ISP in the variational formulation as a change of coordinates, we show here that ISP can equivalently be embedded in that variational formulation by the use of a non-conventional inner-product on the image-space, which allows to work with the image itself instead of a substitute variable. Further, we show that DSP can also be implemented with a non-conventional inner-product on the data-space without any change of coordinate. This way, we restore the symmetry between image-space and data-space, each one having its inner-product that care for its own preconditioning. 

The standard inner products on image-space $X$ ada data-space $Y$ are given by
\begin{equation*}
(a \mid b) = \sum_{i=1}^{\dim(X)} a_i^* b_i, \quad \forall a, b \in X, \quad \text{and} \quad (a \mid b) = \sum_{i=1}^{\dim(Y)} a_i^* b_i, \quad \forall a, b \in Y.
\end{equation*}
These are the inner-products that are conventionally used in MRI reconstruction. In the present article, we also provide $X$ and $Y$ with some non-conventional (generalized) inner-products \((\cdot \mid \cdot)_X\) and \((\cdot \mid \cdot)_Y\) defined respectively as
\begin{equation*}
(a \mid b)_X = \sum_{i,j=1}^{\dim(X)} a_i^* (H_X)_{ij} b_j, \quad (a \mid b)_Y = \sum_{i,j=1}^{\dim(Y)} a_i^* (H_Y)_{ij} b_j,
\end{equation*}
where $H_X$ and $H_Y$ can be any Hermitian positive-definite matrices chosen as desired. These inner-products induce the corresponding $2$-norms
\begin{equation*}
\| x \|_{X,2}^2 = (x \mid x)_X, \quad \| y \|_{Y,2}^2 = (y \mid y)_Y.
\end{equation*}
As we will see, the freedom in the choice of the non-conventional inner-products allows to embed ISP and DSP in the variational formulation of the MRI reconstruction problem withtout change of coordinates by writting it in the form
\begin{equation*}
\text{Find }x^\# \in \underset{x \in X}{\operatorname{argmin}} \frac{1}{2} \| E x - y_0 \|_{Y, 2}^2 + \lambda R(x)
\end{equation*}
\begin{equation*}
\text{with} \quad (\cdot \mid \cdot)_X \quad \text{on} \quad X
\end{equation*}

We finally note that, although ISP is the focus of the present article, DSP also benefits from the introduction of non-conventional inner-products. In fact, DSP has until now merely be considered as a computational heuristic trick to increase convergence speed to the cost of increasing noise and modifying the set of minimizers of the variational problem\cite{ong2020}. We show here that DSP is actually the metric on data-space (just like ISP is the metric on the image-space). It is therefore a fundamental component of the mathematical framework and not just a heuristic. This change of status of preconditioning, from heuristics to formal components, is also an original contribution of our article. 

\subsection{Outline of the Article}

This article presents the two mentioned complementary strategies to incorporate preconditioning into the variational formulation of the MRI reconstruction problem, with the intention to demonstrate their equivalence. The first strategy applies linear transformations to the image space \(X\) and the data-space \(Y\), allowing the reconstruction problem to be expressed using standard Hermitian inner products and the associated \(\ell_2\)-norms. The second strategy avoids any coordinate transformation and instead defines the reconstruction in the original spaces \(X\) and \(Y\), but with non-conventional (generalized) inner-products. We will argue that the first approach encompasses the current state of the art in MRI reconstruction, while the second offers a more intrinsic and theoretically robust formulation—one in which preconditioning becomes an integral part of the underlying mathematical structure.

A preliminary version of this perspective was introduced by the present author in \cite{milaniTTV}, using specific choices for \(H_X\) and \(H_Y\). In the present work, we extend that analysis by showing that the flexibility in choosing \(H_X\) and \(H_Y\) allows embedding both image-space preconditioning (ISP) and data-space preconditioning (DSP) into the variational framework. The structure of the paper is as follows:
\begin{enumerate}
  \item We first revisit the original iterative-SENSE equation, which implicitly contains both ISP (via image intensity correction) and DSP (via $k$-space density compensation). We show that this equation can be reinterpreted either as a problem with coordinate changes and standard inner products, or as one with non-standard inner products but no coordinate changes. Each preconditioning operation appears naturally in one formulation or the other.
  \item We then generalize this equivalence beyond iterative-SENSE, demonstrating that it stems from the invariance of the least-squares problem under isometric transformations.
  \item Once the general mathematical framework has been established in point~$2$, we will return to the normal equations of iterative-SENSE to construct the generalized case: image intensity correction will be extended to general image-space preconditioning (ISP), and $k$-space density compensation will be extended to data-space preconditioning (DSP).
  \item Finally, we incorporate regularization into both formulations, yielding two alternative variational problems: one with coordinate changes and standard inner products, and one with non-standard inner products and unchanged coordinates.
\end{enumerate}

This general framework enables a principled integration of ISP into modern iterative reconstruction methods, including compressed sensing and many iterative deep learning approaches, where it has so far been largely absent \cite{noPrecond1, noPrecond2, noPrecond3, noPrecond4}. By considering preconditioning as a geometric element of the variational model, this work aims to resolve a long-standing gap and provide a unified basis for future algorithmic developments in MRI.

\section{Theory}

\subsection{The Normal Equation and the Conjugate Gradient Descent with Non-Conventional Inner-Products}

The principle of iterative-SENSE is to formulate the reconstruction problem for parallel imaging as a normal equation and then solve it iteratively with the conjugate-gradient descent algorithm (CGD) of Stiefel and Hestenes \cite{CGD}. In this section, we will describe some mathematical facts that pertain to the normal equation and CGD, and which are independent of MRI. In the next section, we will then apply those mathematical facts to the iterative-SENSE reconstruction.

The normal equation is the linear equation
\begin{equation}
A^{\dagger} A u = A^{\dagger} v \tag{$A1$} \label{eq:A1}
\end{equation}
or said differently, we will qualify as "normal equation" any equation of this form. In that equation, $A$ is a linear map from a vector space $U \simeq \mathbb{C}^N$ to another vector space $V\simeq \mathbb{C}^M$, the map $A^{\dagger}$ is its adjoint map from $V$ to $U$, vector $u \in U$ is the unknown of the problem, and vector $v \in V$ is some known data vector. The existence of an adjoint map $A^{\dagger}$ supposes the existence of inner-products on $U$ and $V$. We will write $(\cdot \mid \cdot)_U$ the inner-product on $U$ and $(\cdot \mid \cdot)_V$ the inner-product on $V$. The adjoint $A^{\dagger}$ is then uniquely defined by the relation
\begin{equation*}
(v \mid A u)_V = (A^{\dagger} v \mid u)_U \quad \forall u \in U, \forall v \in V 
\end{equation*}
Notably, the adjoint depends on the inner-product involved, but it always exists. In contrast, the inverse $A^{-1}$ does not exist in general, but it does not depend on the inner-products if it exist. We observe the symmetry
\begin{equation*}
A : U \to V \quad \text{and} \quad A^{\dagger} : V \to U
\end{equation*}
Note that even in the case $A^{-1}$ exists, it is in general different from $A^{\dagger}$. 

If a vector basis is chosen on $U$ and another on $V$, each vector can be written as a coordinate vector, each linear map can be written as a matrix, and each inner-product can be expressed by an inner-product matrix. After some choice of bases, and assuming the field of the vector spaces to be the complex numbers, we may then write $u \in \mathbb{C}^N, v \in \mathbb{C}^M, A \in \mathbb{C}^{M \times N}$ and $A^{\dagger} \in \mathbb{C}^{N \times M}$. Matrix $A$ will be called the "problem matrix" of the normal equation \ref{eq:A1}. Further, we will write $H_U \in \mathbb{C}^{N \times N}$ for the matrix of $(\cdot \mid \cdot)_U$ and we will write $H_V \in \mathbb{C}^{M \times M}$ for the matrix of $(\cdot \mid \cdot)_V$. Both $H_U$ and $H_V$ are then Hermitian positive-definite, and the inner-products can then be written in terms of vectors and matrices as
\begin{equation*}
(u_1 \mid u_2)_U = u_1^* H_U u_2 \quad \forall u_1, u_2 \in U \quad \text{and} \quad (v_1 \mid v_2)_V = v_1^* H_V v_2 \quad \forall v_1, v_2 \in V
\end{equation*}
where $u^*$ stands for the complex conjugate transpose of vector $u$. An important fact is the matrix of the adjoint map verifies
\begin{equation*}
A^{\dagger} = H_U^{-1} A^* H_V
\end{equation*}
where $A^*$ is the complex conjugate transpose of matrix $A$.

The CGD algorithm is a general iterative method to solve the normal equation \ref{eq:A1}. (To be precise, there are two variants of the CGD algorithm presented in the original article \cite{CGD} and the second variant is dedicated to the normal equation). An initial guess $u_0$ is given as argument to the method, and a solution $u^{\#}$ to \ref{eq:A1} is returned by the method after at most $N$ iterations. In the CGD, a notion of conjugacy is defined and depends on the involved inner-products. In the CGD for the normal equation, two vectors $u_1$ and $u_2$ in $U$ (for example two descent directions) are $A^{\dagger} A$ conjugated if
\begin{equation*}
(u_1 \mid A^{\dagger} A u_2)_U = 0
\end{equation*}
or equivalently
\begin{equation*}
(A u_1 \mid A u_2)_V = 0
\end{equation*}
Of note, the CGD as published in its original article assumed that the inner-products on $U$ and $V$ are the standard Hermitian products. In that specific case, the inner-product of two vectors $u_1, u_2 \in U$ is given by
\begin{equation*}
(u_1 \mid u_2)_U = \sum_{i=1}^{N} \overline{u_{1,i}} u_{2,i} = u_1^* u_2
\end{equation*}
and the inner-product of two vectors $v_1, v_2 \in V$ is given by
\begin{equation*}
(v_1 \mid v_2)_V = \sum_{i=1}^{M} \overline{v_{1,i}} v_{2,i} = v_1^* v_2
\end{equation*}
In this specific case, the inner-product matrices $H_U$ and $H_V$ are identity matrices, the matrix of the adjoint map reduces to
\begin{equation*}
A^{\dagger} = A^*
\end{equation*}
and the normal equation \ref{eq:A1} becomes
\begin{equation}
A^* A u = A^* v \tag{$A2$} \label{eq:A2}
\end{equation}
However, all the derivation in the original CGD article can be rewritten in terms of general inner-products and in terms of the adjoint $A^{\dagger}$ instead of $A^*$. The CGD therefore not only solves equation \ref{eq:A2} in the framework of standard Hermitian products, but also solves the more general normal equation \ref{eq:A1} in the framework of non-conventional inner-products. For that, the only adaptations needed in the CGD algorithm are to replace the standard Hermitian products on $U$ resp. $V$ by $(\cdot \mid \cdot)_U$ resp. $(\cdot \mid \cdot)_V$, and to use the adjoint $A^{\dagger}$ instead of the conjugate transpose $A^*$.

At this point, we do two remarks about CGD:
\begin{enumerate}
    \item The original CGD article of Stiefel and Hestenes \cite{CGD} presents an algorithm solving for the unknown $u$ the equation
    \begin{equation}
    A u = v \tag{A3} \label{eq:A3}
    \end{equation}
    and another related but slightly different algorithm solving for the unknown $u$ the normal equation
    \begin{equation}
    A^{\dagger} A u = A^{\dagger} v \tag{A1}
    \end{equation}
    It is therefore the second algorithm that we name CGD in the present article. As mentioned in \cite{CGD}, this second method is equivalent to the first with $A$ being substituted by $A^{\dagger} A$ and $u$ being substituted by $A^{\dagger} u$, but this second algorithm has the advantage of being numerically more efficient.

    \item It is usually claimed that the matrix $A^{\dagger} A$ of the normal equation must be Hermitian positive-definite (and therefore invertible) for the CGD algorithm to be valid (or not ill-posed). This is a misconception. We show formally in a freely available document \cite{littleWolf} that it is sufficient that the matrix $A^{\dagger} A$ is Hermitian non-negative definite (and therefore $A$ can be arbitrary) for the CGD algorithm to converge to a solution of the normal equation in at most $N$ steps. The algorithm is not more ill-posed for $A^{\dagger} A$ non-invertible than for $A^{\dagger} A$ invertible because the problem can always be restricted to subspaces on which the restriction of $A^{\dagger} A$ is invertible. One drawback of having $A^{\dagger} A$ not invertible is that the solution is not unique anymore and the obtained solution depends on the initial value of the CGD (but that dependence is smooth). Another drawback is that some numerical error can accumulate along the iterations but that error is then a vector in $\ker(A)$ and expresses therefore a drift inside the solution set of the normal equation. The solution obtained by CGD is therefore in the solution set of the normal equation even in the presence of that numerical error drift.
\end{enumerate}

We now apply all those facts to the iterative-SENSE reconstruction.

\subsection{The Normal Equation for Iterative-SENSE and the \newline  Preconditioning by a Linear Change of Coordinates}

We will write $E$ the problem matrix (or linear model) of the reconstruction problem as in the original article \cite{iterSense} ($E$ stands for "encoding operator"). In fact, $E$ is the composition of the coil-sensitivity expansion $ C $ and the (uniform or non-uniform) discrete Fourier transform $ F $:
\begin{equation*}
E = F C
\end{equation*}
In order to simplify the present article, we will assume either that k-space noise decorrelation has already been performed prior to any treatment, or that noise is not correlated between coils or sample points. In both cases, the noise-correlation matrix (written $ \tilde{\Psi} $ in the original article) is reduced to the identity. 

The iterative-SENSE article states the reconstruction problem (after noise decorrelation) two times: once without k-space density compensation nor image intensity correction, and a second time with both. In both cases, the reconstruction problem is stated as a normal equation (or an equation that can be rewritten as a normal equation). The authors mention that k-space density compensation and/or image intensity correction can be dropped out or maintained in the problem formulation depending on the preferences of the implementer. They both serve as preconditioners in the sense that they accelerate the convergence of the iterative algorithm for solving it. We will therefore refer to "k-space density-compensation" as a special case of DSP and to "image intensity correction" as a special case of ISP in the present article. Of note, ISP only serves as an accelerator and does not modify the solution set of the reconstruction problem. In contrast, DSP does change the solution set. For that reason, it may not be called a preconditioning by some authors. 

The first statement of the iterative-SENSE reconstruction problem is eq. 13 in \cite{iterSense}, which is a normal equation and which we write here as
\begin{equation}
E^* E x = E^* y_0 \tag{$E2$} \label{eq:E2}
\end{equation}
where $y_0$ is the raw data vector and $x$ is the unknown image to reconstruct. We will write $Y \simeq \mathbb{C}^{nSamp} $ as the vector space containing $y$, and we will write $X \simeq \mathbb{C}^{nVox} $ as the vector space containing $x$. The number of points in the sampling trajectory will be written $nPt$ (for "number of points") and the number of channels (or coils) will be written $nCh$ (for "number of channels") so that
\begin{equation*}
nSamp = nCh \cdot nPt
\end{equation*}
The image vector $x$ is the vertical concatenation of all voxel values. The number $nVox$ is therefore the number of voxels. For a 2D image with $ nVox_x $ rows and $ nVox_y $ columns, it holds then
\begin{equation*}
nVox = nVox_x \cdot nVox_y
\end{equation*}
and this can be extended naturally to 3D images. The matrix $E$ is of size $nSamp \times nVox$. Of note, the original article uses the notation $v$ instead of $x$, $m$ instead of $y$, and $E^H$ instead of $E^*$. In equation \ref{eq:E2}, the adjoint $ E^\dagger $ is equal to $ E^* $, and the inner-products on $ X $ and $ Y $ are assumed to be the standard ones.

\begin{figure} 
    \centering
    \includegraphics[width=0.7\linewidth]{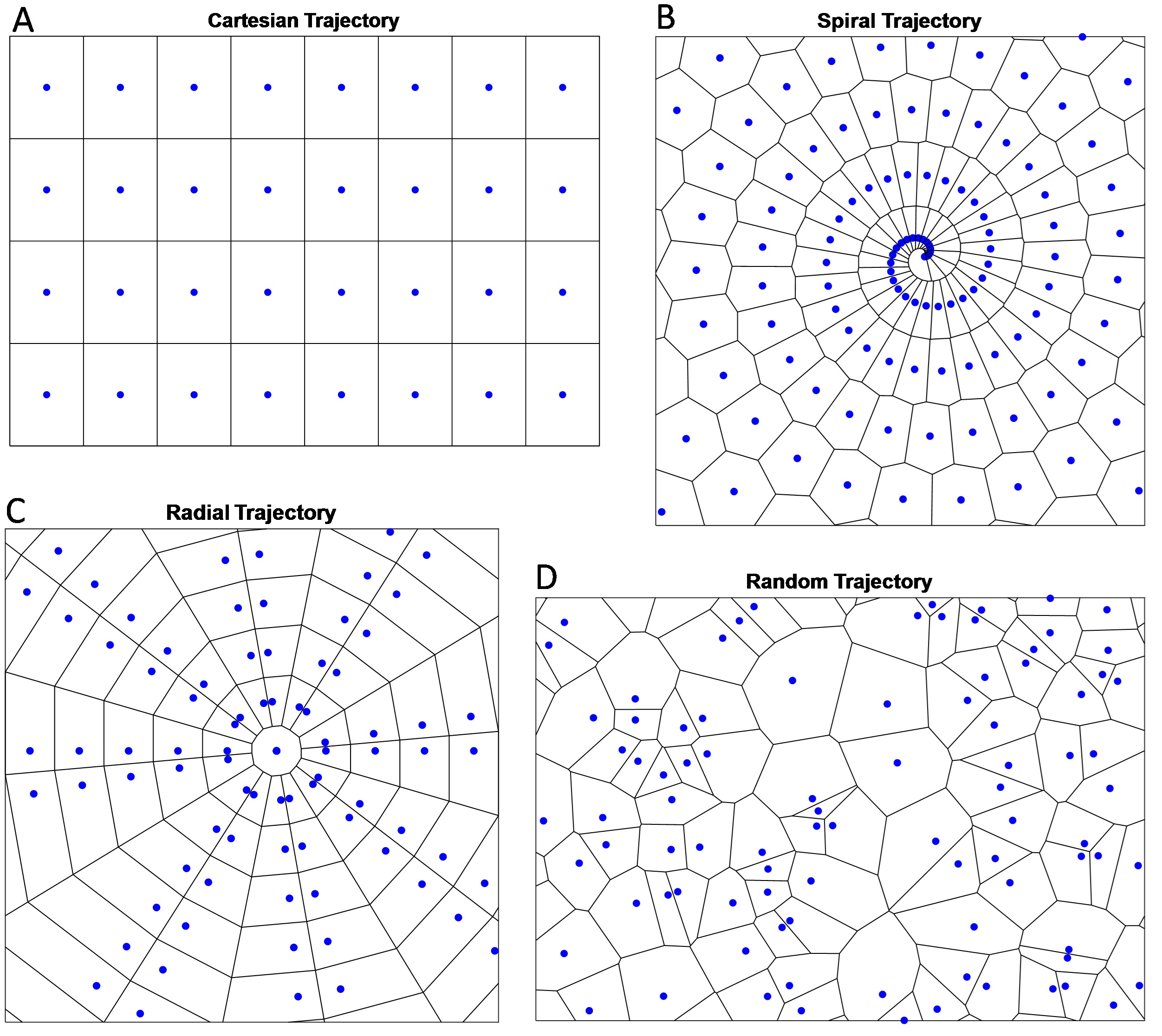}
    \caption{\textit{This figure show four different kinds of trajectories depicted by blue points, together with the Voronoi regions of each point delimited by black edges: a Cartesian trajectory (A), a spiral trajectory (B), a radial trajectory (C), and a random trajectory (D). }}
    \label{fig:figure_ve}
\end{figure}

While the first statement of the reconstruction problem in \cite{iterSense} does not include data-space nor image space preconditioning, the second statement does. It is eq. 24 in \cite{iterSense} and we rewrite it here
\begin{equation}
\left( I E^* D E I \right) \left( I^{-1} x \right) = I E^* D y \tag{$E3$} \label{eq:E3}
\end{equation}
Here, $ D $ is a real-valued diagonal (therefore square) matrix of size $ nSamp \times nSamp $ with positive diagonal elements. It is made of $ nCh \times nCh $ square blocks of size $ nPt \times nPt $, where all diagonal blocks are identical and are equal to a real-valued diagonal matrix with diagonal element number $ \kappa $ equal to 1 divided by the relative density $ d(\cdot) $ in k-space evaluated at trajectory point $ k_\kappa $. Matrix $ D $ is given component-wise in \cite{iterSense} by eq. 20 that we rewrite here again:
\begin{equation*}
D_{\left( \gamma, \kappa \right), \left( \gamma, \kappa \right)} = \frac{1}{d(k_\kappa)}
\end{equation*}
$ D $ is usually referred to in the literature as "(k-space) density compensation". In \cite{iterSense}, it plays the role of the matrix for data-space preconditioning. We can say that density compensation is one special case of data-space preconditioning. It cares for a correct estimation of the functional $\ell_2$-norm of any square-integrable function defined on $k$-space. Given such a function $f_l(\cdot)$ that assigns a complex value to each position $\vec{k}$ of k-space, its functional $\ell_2$-norm can be approximated as
\begin{equation*}
\lVert f_l(\cdot) \rVert_2^2 = \int_{\mathbb{R}^3} dk^3 \ \vert f_l\left(\vec{k}\right) \vert^2 \approx \sum_{m = 1}^{nPt} \Delta K_m \  \vert f_l\left(\vec{k_m}\right) \vert^2 =  \sum_m \Delta K_m \  \vert f_{l, m} \vert^2
\end{equation*}
where $\vec{k}_1, ..., \vec{k}_{nPt}$ is the list of trajectory points, $f_{l, m}$ stands for $f_l\left(\vec{k_m}\right)$ and $\Delta K_m$ is any reasonably chosen positive number to express the volume that points $\vec{k}_m$ occupies. A vector $f$ of data-space can be constructed from $nCh$ functions $f_1(\cdot), ..., f_{nCh}(\cdot)$. The vertical catenation of values $f_{l, 1}, ..., f_{l, nPt}$ of function $f_l(\cdot)$ evaluated on k-space positions $\vec{k}_1, ..., \vec{k}_{nPt}$ forms the vector $f_l$. Further, the vertical catenation of vectors $f_1, ..., f_{nCh}$ is then an element of data-space $f \in \mathbb{C}^{nCh \times nPt}$. Then it makes sense to define its 2-norm as the sum
\begin{equation*}
    \sum_{l = 1}^{nCh} \sum_{m = 1}^{nPt} \Delta K_m \  \vert f_{l, m} \vert^2
\end{equation*}
Typically, one can choose $\Delta K_m$ to be the volume of the Voronoi region of point $\vec{k}_m$ as examplified for different kinds of trajectories in figure \ref{fig:figure_ve}.  
Since the local density of points evaluate in $\vec{k}_m$ is well approximated by $1/\Delta K_m$ we can write

\begin{equation*}
D_{\left( \gamma, \kappa \right), \left( \gamma, \kappa \right)} = \frac{1}{d(k_\kappa)} \approx \Delta K_{\kappa}
\end{equation*}

Matrix $ I $ is also a real-valued diagonal matrix with positive diagonal elements. It is of size $ nVox \times nVox $, and the diagonal element number $ \rho $ is given by the inverse square-root of the sum of the squared magnitudes of the coil-sensitivities evaluated at the position of voxel-center number $ \rho $. It is given by equation 22 in \cite{iterSense}, which we rewrite here:
\begin{equation*}
I_{\rho, \rho} = \frac{1}{\sqrt{\sum_{\gamma=1}^{nCh} |s_\gamma(r_\rho)|^2}}
\end{equation*}
where $ s_\gamma(\cdot) $ is the coil-sensitivity map number $ \gamma $. We note that the inverse $ I^{-1} $ is also a real diagonal matrix with positive diagonal elements, which are given by
\begin{equation*}
(I^{-1})_{\rho, \rho} = \sqrt{\sum_{\gamma=1}^{nCh} |s_\gamma(r_\rho)|^2}
\end{equation*}
We will refer to $ I^{-1} $ as the matrix of image-space preconditioning.

Equation \ref{eq:E3} does not have the form of a normal equation, but this can be fixed. First, we note that
\begin{equation*}
D = D^* \quad \text{and} \quad I = I^*
\end{equation*}
because both are real and diagonal. We will write $ \sqrt{D} $ as the diagonal matrix obtained by taking the square root of $ D $ component-wise, so that
\begin{equation*}
D = \sqrt{D} \sqrt{D}
\end{equation*}
It also holds
\begin{equation*}
\sqrt{D} = \left(\sqrt{D}\right)^*
\end{equation*}
Equation \ref{eq:E3} can then be rewritten as
\begin{equation*}
\left( (\sqrt{D} E I)^* (\sqrt{D} E I) \right) (I^{-1} x) = (\sqrt{D} E I)^* \sqrt{D} y
\end{equation*}
By preforming the substitutions
\begin{equation*}
\tilde{x} \coloneqq I^{-1} x \quad \text{and} \quad \tilde{y} \coloneqq \sqrt{D} y \quad \text{and} \quad \tilde{E} \coloneqq \sqrt{D} E I
\end{equation*}
we obtain the normal equation
\begin{equation}
\tilde{E}^* \tilde{E} \tilde{x} = \tilde{E}^* \tilde{y} \tag{$\tilde{E}2$} \label{eq:tildeE2}
\end{equation}
The substitutions leading to that normal equations are actually equivalent to a linear change of coordinates that we formalize as follows. We define the vector space $\tilde{X} \simeq \mathbb{C}^{nVox} $ and the vector space $ \tilde{Y} \simeq \mathbb{C}^{nSamp} $. We define then the maps
\begin{equation*}
I^{-1}: X \to \tilde{X}, \, x \mapsto I^{-1} x =: \tilde{x}
\end{equation*}
and
\begin{equation*}
\sqrt{D}: Y \to \tilde{Y}, \, y \mapsto \sqrt{D} y =: \tilde{y}
\end{equation*}
The map $ \tilde{E} $ is then from $ \tilde{X} $ to $ \tilde{Y} $
\begin{equation*}
\tilde{E}: \tilde{X} \to \tilde{Y}, \, \tilde{x} \mapsto \tilde{E} \tilde{x}
\end{equation*}
and it holds
\begin{equation*}
\tilde{E} = \sqrt{D} E (I^{-1})^{-1}
\end{equation*}
which is exactly the way $ \tilde{E} $ has to be defined in order to make the following diagram commute:

\begin{equation*}
\begin{tikzcd}
X \arrow{r}{E} \arrow[swap]{d}{I^{-1}}  &  Y \arrow{d}{\sqrt{D}} \\
\tilde{X} \arrow{r}{\tilde{E}}   &    \tilde{Y} 
\end{tikzcd}
\end{equation*}

We note that k-space and image-space preconditioning can be dropped out by setting $ D $ or $ I $ equal to the identity. If both are dropped out, equation \ref{eq:tildeE2} simplifies to equation \ref{eq:E2}. Since equation \ref{eq:E2} is a special case of \ref{eq:tildeE2}, we will refer to \ref{eq:tildeE2} as the normal equation of iterative-SENSE.

Performing iterative-SENSE consists of solving the normal equation \ref{eq:tildeE2} for $ \tilde{x} $ with the CGD algorithm. The final image $ x $ can then be recovered as a final step by multiplying the solution $ \tilde{x} $ by $ I $. Of note, and to be fair from a historical point of view, the method published in the original article \cite{iterSense} solves equation \ref{eq:E3} with problem matrix $ I E^* D E I $ and with the CGD algorithm for equation \ref{eq:A3}, i.e., not the CGD for the normal equation. It is however mathematically equivalent to our formulation. We can say that it is a different implementation of the same method.

Whatever CGD algorithm is used for the iterative-SENSE method, the notion of conjugacy used in that case is dictated by the standard inner-products, and the adjoint of $ \tilde{E} $ is given by $ \tilde{E}^* $ in accordance with the theory explained above. For example, it means that the inner-product of two vectors $ \tilde{x}_1 $ and $ \tilde{x}_2 $ is equal to $ \tilde{x}_1^* \tilde{x}_2 $. However, the inner-product of two vectors $ x_1 $ and $ x_2 $ is not equal to $ x_1^* x_2 $. This apparent contradiction gets resolved in the next section, where we explain that $ x $ and $ \tilde{x} $ are two different manifestation of the same vector, but that those manifestations live in two different vector spaces with their own notion of inner-product.

\subsection{Another Normal Equation for Iterative-SENSE and the \newline  Preconditioning by Non-Conventional Inner-Products}

On $ \tilde{X} $, we define the inner-product $ (\cdot | \cdot)_{\tilde{X}} $ by
\begin{equation*}
(\tilde{x}_1 | \tilde{x}_2)_{\tilde{X}} \coloneqq \tilde{x}_1^* H_{\tilde{X}} \tilde{x}_2 = \tilde{x}_1^* \tilde{x}_2 \quad \forall \, \tilde{x}_1, \tilde{x}_2 \in \tilde{X}
\end{equation*}
meaning that the inner-product matrix $ H_{\tilde{X}} $ is set equal to the identity. We do the same on $ \tilde{Y} $:
\begin{equation*}
(\tilde{y}_1 | \tilde{y}_2)_{\tilde{Y}} \coloneqq \tilde{y}_1^* H_{\tilde{Y}} \tilde{y}_2 = \tilde{y}_1^* \tilde{y}_2 \quad \forall \, \tilde{y}_1, \tilde{y}_2 \in \tilde{Y}
\end{equation*}
We observe now
\begin{equation*}
(\tilde{x}_1 | \tilde{x}_2)_{\tilde{X}} = \tilde{x}_1^* \tilde{x}_2 = x_1^* I^{-1} I^{-1} x_2 = x_1^* (I^{-1})^2 x_2
\end{equation*}
and
\begin{equation*}
(\tilde{y}_1 | \tilde{y}_2)_{\tilde{Y}} = \tilde{y}_1^* \tilde{y}_2 = y_1^* \sqrt{D} \sqrt{D} y_2 = y_1^* D y_2
\end{equation*}
We define
\begin{equation}
H_X \coloneqq (I^{-1})^2 \quad \text{and} \quad H_Y \coloneqq D \tag{$DI1$} \label{eq:DI1}
\end{equation}
in order to define the inner-products on $ X $ and $ Y $ as
\begin{equation*}
(x_1 | x_2)_X \coloneqq x_1^* H_X x_2 \quad \text{and} \quad (y_1 | y_2)_Y \coloneqq y_1^* H_Y y_2
\end{equation*}
We have this way
\begin{equation*}
(x_1 | x_2)_X = (\tilde{x}_1 | \tilde{x}_2)_{\tilde{X}} \quad \text{and} \quad (y_1 | y_2)_Y = (\tilde{y}_1 | \tilde{y}_2)_{\tilde{Y}}
\end{equation*}
so that the maps $ I^{-1} $ and $ \sqrt{D} $ performing the change of coordinates become isometries of inner-product spaces. The adjoint of $ E $ with respect to those inner-products is given by
\begin{equation*}
E^\dagger = H_X^{-1} E^* H_Y = I^2 E^* D
\end{equation*}
Equation \ref{eq:E3}, which is equivalent to \ref{eq:tildeE2}, can be rewritten as:  
\begin{equation*}
I E^* D E x = I E^* D y
\end{equation*}
Multiplying both sides by the left with $ I $ leads to
\begin{equation*}
I^2 E^* D E x = I^2 E^* D y
\end{equation*}
which is an equation already established by in \cite{iterSense} (eq. 23) but which is nothing else than the normal equation
\begin{equation}
E^\dagger E x = E^\dagger y  \tag{$E1$} \label{eq:E1}
\end{equation}
in the coordinates with non-conventional inner-products. We have so far defined each quantity in two "worlds". In the first world, $ x $ is represented by $ \tilde{x} $, $ E $ is represented by $ \tilde{E} $, $ (\cdot | \cdot)_X $ is represented by $ (\cdot | \cdot)_{\tilde{X}} $, and so on. Because the symbol "$\ \tilde{\ } \ $" is called a "tilde", we will say that any quantity written with a tilde is written in the "tilde coordinates". In the second world, $x$ is represented by $x$ itself, $E$ is represented by $E$ itself, and so on. We will say that any quantity written without a tilde is written in the "direct coordinates". Every symbol represented in the direct coordinates can be written in the tilde coordinates and vice versa. 

Equation \ref{eq:E1} is the normal equation of iterative-SENSE in the direct coordinates where the inner-products $ (\cdot | \cdot)_X $ and $ (\cdot | \cdot)_Y $ are non-conventional. While the inner-product matrix of $ (\cdot | \cdot)_X $ is the manifestation of image-space preconditioning (ISP), the inner-product matrix of $ (\cdot | \cdot)_Y $ is the manifestation of data-space preconditioning (DSP). As we said in the first theory section, normal equation \ref{eq:E1} can be solved with CGD using the non-conventional inner-products $ (\cdot | \cdot)_X $ and $ (\cdot | \cdot)_Y $ and using $ E^\dagger $ for the adjoint map (and not $ E^* $ as given in the original CGD publication). This offers an alternative and fully equivalent method to the original iterative-SENSE reconstruction where \ref{eq:tildeE2} is solved with the CGD considering standard Hermitian products and the adjoint map $ \tilde{E}^* $. We can say that the original iterative-SENSE method operates in the tilde coordinates, while solving equation \ref{eq:E1} operates in the direct coordinates.

\subsection{The Least-Square Problem Associated with \newline the Normal Equation}

Again, we present in this section purely mathematical facts that are independent of MRI. We will then apply them to iterative-SENSE in another section hereafter.

Given a vector $ v \in V $ and a matrix $ A $ which maps any vector $ u \in U $ into $ V $, we recall equation \ref{eq:A1}, which is the normal equation associated with $ A $ and $ v $:

\begin{equation}
A^{\dagger} A u = A^{\dagger} v \tag{$A1$}
\end{equation}

We recall that the notion of adjoint implies tacitly the presence of the inner-products $ (\cdot | \cdot)_U $ on $ U $ and $ (\cdot | \cdot)_V $ on $ V $. These inner-products naturally induce a 2-norm on their respective spaces, namely:

\begin{equation*}
\| u \|_{U,2}^2 = (u | u)_U \quad \forall u \in U \quad \text{and} \quad \| v \|_{V,2}^2 = (v | v)_V \quad \forall v \in V
\end{equation*}

In the following, we will write $ Ker(A)$ for the kernel of $A$, i.e., the subspace of $U$ given by

\begin{equation*}
Ker(A) \coloneqq \{ u \in U | A u = 0 \}
\end{equation*}
We will write $ S_A $ the solution set to equation \ref{eq:A1}. It can be shown that $ S_A $ is never empty, but it may contain more than one solution. The solution is unique exactly if $ Ker(A) = \{0\} $, in which case $ S_A $ is a singleton. If $ Ker(A) \neq \{0\} $, then $ S_A $ is an affine space parallel to $ Ker(A) $.
It can be shown that the set $ S_A $ is equal to the argmin-set of the squared-norm function
\begin{equation}
u \mapsto \frac{1}{2} \| A u - v \|_{V,2}^2 \tag{$FA$} \label{eq:FA}
\end{equation}
We write therefore
\begin{equation*}
S_A = \underset{u \in U}{\operatorname{argmin}} \frac{1}{2} \| A u - v \|_{V,2}^2 
\end{equation*}
The squared-norm function is the objective of the optimization problem
\begin{equation*}
\text{Find } u^{\#} \in S_A = \underset{u \in U}{\operatorname{argmin}} \frac{1}{2} \| A u - v \|_{V,2}^2 \tag{PA} \label{eq:PA}
\end{equation*}
The set of minimizers $S_A$ of that optimization problem is the solution set of the normal equation \ref{eq:A1}. We will say that normal equation \ref{eq:A1} and problem \ref{eq:PA} are associated to each other. We have the equivalence

\begin{equation*}
A^{\dagger} A u^{\#} = A^{\dagger} v \quad \Longleftrightarrow \quad u^{\#} \in S_A := \underset{u \in U}{\operatorname{argmin}} \frac{1}{2} \| A u - v \|_{V,2}^2
\end{equation*}

Finding a solution $u^{\#}$ of \ref{eq:PA} can therefore be achieved by the CGD algorithm with any initial value $ u_0 $. If $Ker(A) \neq \{0\}$, then there exist many solutions and $u^{\#}$ depends on the initial value $ u_0 $. As we show in \cite{littleWolf}, the CGD performs the orthogonal projection of the initial value $ u_0 $ onto the solution set $S_A $.

Again, we want to stress the fact that the CGD algorithm remains valid even if $A^{\dagger} A$ is not positive-definite (i.e., not invertible, since $ A^{\dagger} A $ is Hermitian non-negative definite in any case). We invite the reader to consult \cite{littleWolf} for a formal proof. We also recall that solving \ref{eq:A1} with CGD implies using the (possibly non-conventional) inner-products $(\cdot | \cdot)_U$ on $ U $ and $ (\cdot | \cdot)_V $ on $ V $ for conjugacy and to use accordingly the adjoint $A^{\dagger}$ instead of the conjugate transpose $A^*$.

For completeness, we terminate this section by introducing the notion of differentiation we need in MRI reconstruction and we use it to demonstrate the equivalence of \ref{eq:A1} and \ref{eq:PA}. Let

\begin{equation*}
f(\cdot) : U \to \mathbb{R}, \quad u \mapsto f(u) \in \mathbb{R}
\end{equation*}
be a real-valued function. We will write the real and imaginary part of $ u $ as

\begin{equation*}
r u := \text{real}(u) \in \mathbb{R}^N, \quad i u := \text{imag}(u) \in \mathbb{R}^N
\end{equation*}
which are notably two real vectors. Writing the square root of -1 as $ j $, it follows
\begin{equation*}
u = r u + j \  i u
\end{equation*}
We will say that function $ f(\cdot) $ is differentiable if it is differentiable in the sense of real differentiation with respect to both independent variables $ r u $ and $ i u $. We define the derivative $ \frac{\partial}{\partial u} $ acting on any differentiable function $ f(\cdot) $ by

\begin{equation*}
\frac{\partial f(u)}{\partial u} := \frac{\partial f(u)}{\partial r u} + j \frac{\partial f(u)}{\partial i u}
\end{equation*}

On the other hand, we will write the gradient of function $ f(\cdot) $ evaluated at position $ u $ as $ grad_u f \in U $, which is a vector in $ U $. We define this gradient as the only vector in $ U $ satisfying

\begin{equation*}
(grad_u f | h)_U = \left. \frac{d}{dt} f(u + t \cdot h) \right|_{t=0} \quad \forall h \in U
\end{equation*}

The gradient defined this way always exists and is always unique \cite{littleWolf}. As the reader may have noticed, the definition of $ \frac{\partial f(u)}{\partial u} $ is independent of $ H_U $ while the definition of $ grad_u f $ depends on $ H_U $. A straightforward calculation shows that both are related by
\begin{equation*}
grad_u f = H_U^{-1} \frac{\partial f(u)}{\partial u}
\end{equation*}
It can then be shown that the gradient of function \ref{eq:FA} verifies
\begin{equation*}
grad_u \frac{1}{2} \| A u - v \|_{V,2}^2 = H_U^{-1} \frac{\partial}{\partial u} \frac{1}{2} \| A u - v \|_{V,2}^2 = H_U^{-1} A^* H_V (A u - v)
\end{equation*}
The substitution of the expression for the adjoint leads to
\begin{equation*}
grad_u \frac{1}{2} \| A u - v \|_{V,2}^2 = A^{\dagger} (A u - v)
\end{equation*}
Since the objective function is differentiable and convex, a necessary and sufficient condition for $ u $ to be a minimizer is
\begin{equation*}
grad_u \frac{1}{2} \| A u - v \|_{V,2}^2 = 0
\end{equation*}
By substitution of the expression above for the gradient of \ref{eq:FA} leads to the equation
\begin{equation*}
A^{\dagger} (A u - v) = 0
\end{equation*}
and after reformatting
\begin{equation*}
A^{\dagger} A u = A^{\dagger} v
\end{equation*}
which is nothing else than the normal equation \ref{eq:A1}. It demonstrates the equivalence between \ref{eq:A1} and \ref{eq:PA}.

\subsection{The Invariance of the Least-Square Problem \newline under an Isometrical Change of Coordinates}
In this section, the change of coordinates introduced for iterative-SENSE is generalized to any isometrical change of coordinates between finite dimensional complex vector spaces with arbitrary inner-products. The corresponding change of coordinates for the associated least-square problem follows naturally. 

Let $\varphi$ be an isomorphism (an invertible linear map) from the vector space $U$ to another vector space $\tilde{U}$. For any $u \in U$ we define 
\begin{equation*}
\tilde{u} \coloneqq \varphi u.
\end{equation*}
We have thus
\begin{equation*}
\varphi: U \to \tilde{U}, \quad u \mapsto \varphi u = \tilde{u}.
\end{equation*}
Further, let $\psi$ be an isomorphism from the vector space $V$ to another vector space $\tilde{V}$ and we define
\begin{equation*}
\tilde{v} \coloneqq \psi v
\end{equation*}
for any $v \in V$. therefore
\begin{equation*}
\psi: V \to \tilde{V}, \quad v \mapsto \psi v = \tilde{v}.
\end{equation*}
These definitions are written in a way that suggests that $u$ can be chosen, and $\tilde{u}$ follows from the choice of $u$. But it can also be interpreted the other way around. We can choose a vector $\tilde{u} \in \tilde{U}$ and deduce its pre-image $u \in U$ by the map $\varphi$. The vectors of $U$ and $\tilde{U}$ are paired by the isomorphism $\varphi$. A vector can be chosen either in $U$ or $\tilde{U}$, and its counterpart in the other space follows. The same holds between $V$ and $\tilde{V}$.

Let $(\cdot | \cdot)_U$ be an inner-product on $U$ and $(\cdot | \cdot)_V$ be an inner-product on $V$. The induced 2-norms on $U$ and $V$ are then given by
\begin{equation*}
\|u\|_{U,2}^2 = (u | u)_U \quad \forall u \in U, \quad \text{resp.} \quad \|v\|_{V,2}^2 = (v | v)_V \quad \forall v \in V.
\end{equation*}
We define then on $\tilde{U}$ and $\tilde{V}$ the inner-products $(\cdot | \cdot)_{\tilde{U}}$ and $(\cdot | \cdot)_{\tilde{V}}$ by
\begin{equation}
(\tilde{u}_1 | \tilde{u}_2)_{\tilde{U}} \coloneqq (u_1 | u_2)_U \quad \forall \tilde{u}_1, \tilde{u}_2 \in \tilde{U}, \quad \text{resp.} \quad (\tilde{v}_1 | \tilde{v}_2)_{\tilde{V}} \coloneqq (v_1 | v_2)_V \quad \forall \tilde{v}_1, \tilde{v}_2 \in \tilde{V} \tag{$H1$} \label{eq:H1}
\end{equation}
They induce the 2-norms on $\tilde{U}$ and $\tilde{V}$ as given by
\begin{equation*}
\|\tilde{u}\|_{\tilde{U},2}^2 = (\tilde{u} | \tilde{u})_{\tilde{U}} \quad \forall \tilde{u} \in \tilde{U}, \quad \text{resp.} \quad \|\tilde{v}\|_{\tilde{V},2}^2 = (\tilde{v} | \tilde{v})_{\tilde{V}} \quad \forall \tilde{v} \in \tilde{V}.
\end{equation*}
It follows in particular that $\varphi$ and $\psi$ are isometries of inner-product spaces. We observe therefore for any $\tilde{u}_1, \tilde{u}_2 \in \tilde{U}$
\begin{equation*}
(\tilde{u}_1 | \tilde{u}_2)_{\tilde{U}} = (\varphi u_1 | \varphi u_2)_{\tilde{U}} = (u_1 | u_2)_U.
\end{equation*}
It follows
\begin{equation*}
(u_1 | \varphi^\dagger \varphi u_2)_U = (u_1 | u_2)_U \quad \forall u_1, u_2 \in U,
\end{equation*}
and the same holds for $\psi$ between $V$ and $\tilde{V}$. It holds thus
\begin{equation*}
\varphi^{-1} = \varphi^\dagger \quad \text{and} \quad \psi^{-1} = \psi^\dagger.
\end{equation*}
which means that both $\varphi$ and $\psi$ are unitary. This is a consequence of our choice of inner-products on $\tilde{U}$ and $\tilde{V}$. We also note that instead of defining $(\cdot | \cdot)_{\tilde{U}}$ and $(\cdot | \cdot)_{\tilde{V}}$ as a function of $(\cdot | \cdot)_U$ and $(\cdot | \cdot)_V$, we can also do the opposite: given $(\cdot | \cdot)_{\tilde{U}}$ and $(\cdot | \cdot)_{\tilde{V}}$ we can define
\begin{equation}
(u_1 | u_2)_U \coloneqq (\tilde{u}_1 | \tilde{u}_2)_{\tilde{U}} \quad \forall u_1, u_2 \in U, \quad \text{resp.} \quad (v_1 | v_2)_V \coloneqq (\tilde{v}_1 | \tilde{v}_2)_{\tilde{V}} \quad \forall v_1, v_2 \in V \tag{$H2$} \label{eq:H2}
\end{equation}
and we obtain the same results. In any case, the change of coordinates are isometries. 

By the choice of a vector basis on $U$, the inner-product $(\cdot | \cdot)_U$ can be expressed by a Hermitian positive-definite matrix $H_U$ as
\begin{equation*}
(u_1 | u_2)_U = u_1^* H_U u_2.
\end{equation*}
The same holds on $\tilde{U}$:
\begin{equation*}
(\tilde{u}_1 | \tilde{u}_2)_{\tilde{U}} = \tilde{u}_1^* H_{\tilde{U}} \tilde{u}_2.
\end{equation*}
We will read $\varphi$ as a matrix according to the chosen vector bases on $U$ and $\tilde{U}$. By $\varphi^*$ we mean then the complex conjugate transpose of the matrix $\varphi$. Note that $\varphi^*$ and $\varphi^\dagger$ do not coincide in general. By definition of the inner-products, it holds thus
\begin{equation*}
\tilde{u}_1^* H_{\tilde{U}} \tilde{u}_2 = u_1^* \varphi^* H_{\tilde{U}} \varphi u_2 = u_1^* H_U u_2 \quad \forall u_1, u_2 \in U.
\end{equation*}
And therefore
\begin{equation*}
\varphi^* H_{\tilde{U}} \varphi = H_U.
\end{equation*}
All that was said on $U$ and $\tilde{U}$ is also true for $V$ and $\tilde{V}$. By a choice of vector bases, the inner-products $(\cdot | \cdot)_V$ and $(\cdot | \cdot)_{\tilde{V}}$ are thus expressed by Hermitian positive-definite matrices $H_V$ and $H_{\tilde{V}}$ verifying
\begin{equation*}
\psi^* H_{\tilde{V}} \psi = H_V.
\end{equation*}

Now let $A$ be a linear map (a homomorphism) from $U$ to $V$. We define then the linear map $\tilde{A}$ from $\tilde{U}$ to $\tilde{V}$ so that the following diagram commutes:

\begin{equation*}
\begin{tikzcd}
U \arrow{r}{A} \arrow[swap]{d}{\varphi}  &  V \arrow{d}{\psi} \\
\tilde{U} \arrow{r}{\tilde{A}}   &    \tilde{V} 
\end{tikzcd}
\end{equation*}
It means
\begin{equation*}
\tilde{A} = \psi A \varphi^{-1}.
\end{equation*}
One can then check that
\begin{equation*}
\tilde{A}^\dagger = \varphi A^\dagger \psi^{-1}.
\end{equation*}
A straightforward calculation shows then the equivalence between normal equation \ref{eq:A1} and the normal equation
\begin{equation}
\tilde{A}^\dagger \tilde{A} \tilde{u} = \tilde{A}^\dagger \tilde{v} \tag{$\tilde{A}1$} \label{eq:tildeA1}
\end{equation}
By "equivalent" we mean that $\tilde{u}$ is a solution of \ref{eq:tildeA1} if and only if $u$ is a solution of \ref{eq:A1}. If the inner-products in the tilde coordinates are the standard inner-products, equation \ref{eq:A1} becomes
\begin{equation*}
\tilde{A}^* \tilde{A} \tilde{u} = \tilde{A}^* \tilde{v} \tag{$\tilde{A}2$} \label{eq:tildeA2}
\end{equation*}
But we will consider the general case \ref{eq:tildeA1}.

As in the previous section, we will say that the quantities
\begin{equation*}
\tilde{A}, \tilde{U}, \tilde{V}, \tilde{u}, \tilde{v}, (\cdot|\cdot)_{\tilde{U}}, (\cdot|\cdot)_{\tilde{V}}, \dots  
\end{equation*}
are expressed in the tilde coordinates. Further, we will say that the quantities 
\begin{equation*}
A, U, V, u, v, (\cdot|\cdot)_U, (\cdot|\cdot)_V, \dots 
\end{equation*}
are expressed in the direct coordinates. Every quantity that can be expressed in the tilde coordinates can be expressed by a representative in the direct coordinates and reversely. Given a real-valued function $ f $ defined on $ U $, we refer to the previous section for the definition of the gradient $ grad_u f $ and the derivative $ \frac{\partial}{\partial u} $. We define the representant of function $ f $ in the tilde coordinates as the function 
\begin{equation*}
\tilde{f}: \tilde{U} \to \mathbb{R}, \quad \tilde{u} \mapsto \tilde{f}(\tilde{u}) := f(u).
\end{equation*}
In other words,
\begin{equation*}
\tilde{f}(\tilde{u}) = f(\varphi^{-1} \tilde{u}).
\end{equation*}
The gradient transforms as
\begin{equation*}
grad_{\tilde{u}} \tilde{f} = \varphi grad_u f,
\end{equation*}
while the derivative $ \frac{\partial}{\partial u} $ does not transform in a natural linear way. Rather it holds
\begin{equation*}
\frac{\partial}{\partial \tilde{u}} \tilde{f}(\tilde{u}) = (\varphi^{-1})^* \frac{\partial}{\partial u} f(u) = H_{\tilde{U}} \varphi H_U^{-1} \frac{\partial}{\partial u} f(u).
\end{equation*}
Demonstrating those relations can be counter-intuitive since we work here with complex-valued variables while the differentiation is defined in the real sense. In order to help the reader reproduce the proofs, we give a few hints which merely consist in rewriting the formulas in what we call the "real representation" in \cite{littleWolf}. The original representation will be called the "complex representation". 

For the complex vector spaces $ U \simeq \mathbb{C}^N $ and $ \tilde{U} \simeq \mathbb{C}^N $, we write their real representation as $ \mathcal{R} U \simeq \mathbb{R}^{2N} $ and $ \mathcal{R} \tilde{U} \simeq \mathbb{R}^{2N} $. We define the real representation of $ u \in U $ and $ \tilde{u} \in \tilde{U} $ as
\begin{equation*}
\mathcal{R}u = \begin{bmatrix} ru \\ iu \end{bmatrix} \in \mathcal{R}U, \quad \mathcal{R}\tilde{u} = \begin{bmatrix} r\tilde{u} \\ i\tilde{u} \end{bmatrix} \in \mathcal{R}\tilde{U}.
\end{equation*}
We define the real representation of $ \varphi $ as the block matrix
\begin{equation*}
\mathcal{R}\varphi = \begin{bmatrix} r\varphi & -i\varphi \\ i\varphi & r\varphi \end{bmatrix},
\end{equation*}
which is the matrix of a map from $ \mathcal{R}U $ to $ \mathcal{R}\tilde{U} $. Here is $r\varphi$ is the matrix consisting of the real part of $\varphi$ and $i\varphi$ is the matrix consisting of its imaginary part. It can then be checked that
\begin{equation*}
\mathcal{R}(\varphi u) = \mathcal{R}\varphi \mathcal{R}u, \quad  \quad \mathcal{R}(\varphi^{-1}) = \left(\mathcal{R}\varphi\right)^{-1}, \quad \mathcal{R}(\varphi^*) = \left(\mathcal{R}\varphi\right)^T.
\end{equation*}
All equations written in the complex representation can equivalently be written in the real representation. For example, it holds the equivalence
\begin{equation*}
\tilde{u} = \varphi u \quad \Longleftrightarrow \quad \mathcal{R}\tilde{u} = \mathcal{R}\varphi \mathcal{R}u.
\end{equation*}
We define the real representation of the function $ f $ as
\begin{equation*}
\mathcal{R}f: \mathcal{R}U \to \mathbb{R}, \quad \begin{bmatrix} ru \\ iu \end{bmatrix} \mapsto \mathcal{R}f\left(\begin{bmatrix} ru \\ iu \end{bmatrix}\right) \coloneqq f(u).
\end{equation*}
We define the real representation of $ \frac{\partial}{\partial u} $ and $ \frac{\partial}{\partial \tilde{u}} $ as
\begin{equation*}
\mathcal{R}\left(\frac{\partial}{\partial u}\right) = \begin{bmatrix} \frac{\partial}{\partial ru} \\ \frac{\partial}{\partial iu} \end{bmatrix}, \quad \mathcal{R}\left(\frac{\partial}{\partial \tilde{u}}\right) = \begin{bmatrix} \frac{\partial}{\partial r\tilde{u}} \\ \frac{\partial}{\partial i\tilde{u}} \end{bmatrix}.
\end{equation*}
It is then a matter of real differential calculus to show that
\begin{equation*}
\begin{bmatrix} \frac{\partial}{\partial r\tilde{u}} \\ \frac{\partial}{\partial i\tilde{u}} \end{bmatrix} 
\mathcal{R}\tilde{f}\left(\begin{bmatrix} r\tilde{u} \\ i\tilde{u} \end{bmatrix}\right) = 
\begin{bmatrix} \frac{\partial}{\partial r\tilde{u}} \\ \frac{\partial}{\partial i\tilde{u}} \end{bmatrix} 
\mathcal{R}f\left(\mathcal{R}\varphi^{-1} \begin{bmatrix} r\tilde{u} \\ i\tilde{u} \end{bmatrix}\right) = 
(\mathcal{R}\varphi^{-1})^T \begin{bmatrix} \frac{\partial}{\partial ru} \\ \frac{\partial}{\partial iu} \end{bmatrix} 
\mathcal{R}f\left(\begin{bmatrix} ru \\ iu \end{bmatrix}\right)
\end{equation*}
which translates back in the complex representation by
\begin{equation*}
\frac{\partial}{\partial \tilde{u}} \tilde{f}(\tilde{u}) = \frac{\partial}{\partial \tilde{u}} f(\varphi^{-1} \tilde{u}) = (\varphi^{-1})^* \frac{\partial}{\partial u} f(u).
\end{equation*}
Multiplying both sides by $ H_{\tilde{U}}^{-1} $ from the left leads to
\begin{equation*}
H_{\tilde{U}}^{-1} \frac{\partial}{\partial \tilde{u}} \tilde{f}(\tilde{u}) = H_{\tilde{U}}^{-1} (\varphi^{-1})^* H_U H_U^{-1} \frac{\partial}{\partial u} f(u).
\end{equation*}
Noting that
\begin{equation*}
H_{\tilde{U}}^{-1} (\varphi^{-1})^* H_U = (\varphi^{-1})^\dagger \quad \text{and} \quad \varphi^{-1} = \varphi^\dagger,
\end{equation*}
leads to
\begin{equation*}
H_{\tilde{U}}^{-1} \frac{\partial}{\partial \tilde{u}} \tilde{f}(\tilde{u}) = \varphi H_U^{-1} \frac{\partial}{\partial u} f(u),
\end{equation*}
and therefore, we conclude
\begin{equation*}
grad_{\tilde{u}} \tilde{f} = \varphi grad_u f.
\end{equation*}
It means that the gradient $grad_u f$ transforms like any vector in $U$. 

We finally come to the least-squares problem. As seen in the previous section, the normal equation \ref{eq:A1} is equivalent to the least-squares problem \ref{eq:PA}. All what is true in the direct coordinates is also true in the tilde coordinates. We deduce that the normal equation \ref{eq:tildeA1} is equivalent to the least-squares problem 
\begin{equation}
\text{Find } \tilde{u}^\# \in S_{\tilde{A}} = \underset{\tilde{u} \in \tilde{U}}{\operatorname{argmin}} \frac{1}{2} \|\tilde{A} \tilde{u} - \tilde{v} \|_{\tilde{V},2}^2 \tag{$P\tilde{A}$} \label{eq:PtildeA}
\end{equation}

As we have seen, \ref{eq:A1} is equivalent to \ref{eq:tildeA1} in the sense that $u$ is a solution of \ref{eq:A1} if and only if $\tilde{u}$ is a solution of \ref{eq:tildeA1}. It follows that $u$ is a minimizer in problem \ref{eq:PA} exactly if $\tilde{u}$ is a minimizer in problem \ref{eq:PtildeA}. It holds
\begin{equation*}
S_A = \varphi^{-1} S_{\tilde{A}}  := \{\varphi^{-1} \tilde{u} \mid \tilde{u} \in S_{\tilde{A}}\},
\end{equation*}
so that the notation is well-posed.

We have shown in the two previous sections that the normal equation and the associated least-squares problem are invariant under a linear change of coordinates if the inner-products are defined appropriately, i.e., as in definitions \ref{eq:H1} or \ref{eq:H2}. 

As we will see in the next section, the transition to the tilde coordinates is a generalization of the preconditioning introduced for iterative-SENSE, where $ \varphi $ takes the meaning of the image space preconditioning and $ \psi $ takes the meaning of the data-space preconditioning. In the direct coordinates, $ \varphi $ is absorbed in the inner-product matrix $ H_U $, and $ \psi $ is absorbed in the inner-product matrix $ H_V $. therefore, in the direct coordinates, $ H_U $ is the manifestation of image space preconditioning, and $ H_V $ is the manifestation of the data-space preconditioning.

\subsection{Regularized Least-Square Reconstructions}
We now apply the mathematical material presented in the two previous sections to the normal equation of iterative-SENSE and to the associated least-square problem. We will then deduce a formulation of regularized least-square reconstructions in the direct and in the tilde coordinates, so that we will be able to compare the two. We begin by replacing each symbol of the general mathematical framework presented above by a symbol having a meaning in the context of MRI. 

We set $X \simeq \mathbb{C}^{nVox} $ (instead of $U$) to be the vector space containing MRI images and $Y \simeq \mathbb{C}^{nSamp} $ (instead of $V$) to be the vector space containing the raw data. We will write $x \in X$ (instead of $u \in U$) for any MRI image and $y \in Y$ (instead of $v \in V$) for any raw data vector. Let the change of coordinates be given by any two isomorphisms
\begin{equation*}
\varphi \colon X \to \tilde{X}, \quad \psi \colon Y \to \tilde{Y}
\end{equation*}
as in the previous section. Here, $\tilde{X} \simeq \mathbb{C}^{nVox} $ (instead of $\tilde{U} $) is the tilde representative of $X$, and further $\tilde{Y}\simeq \mathbb{C}^{nSamp} $ (instead of $\tilde{V}$) is the tilde representative of $Y$. The tilde representative of $x$ will be written $\tilde{x} \in \tilde{X} $ (instead of $ \tilde{u} \in \tilde{U} $) and the tilde representative of $y$ will be written $\tilde{y} \in \tilde{Y} $ (instead of $\tilde{v} \in \tilde{V}$). We call $H_{\tilde{X}}$ the inner-product matrix on $ \tilde{X} $ (instead of $H_{\tilde{U}}$) and we call $H_{\tilde{Y}}$ the inner-product matrix on $\tilde{Y}$ (instead of $H_{\tilde{V}}$). 

We now restrict the general framework exposed in the previous section to a special case: we set the inner-product matrices $ H_{\tilde{X}} $ and $ H_{\tilde{Y}} $ to be identity matrices. It follows
\begin{equation}
H_X = \varphi^* H_{\tilde{X}} \varphi = \varphi^* \varphi \quad \text{and} \quad H_Y = \psi^* H_{\tilde{Y}} \psi = \psi^* \psi \tag{$H3$} \label{eq:H3}
\end{equation}
where $ H_X $ is the inner-product matrix on $ X $ (instead of $ H_U $) and where $ H_Y $ is the inner-product matrix on $ Y $ (instead of $ H_V $).

As a consequence, the inner-products on $ \tilde{X} $ and $ \tilde{Y} $ are the conventional ones, i.e.
\begin{equation*}
(\tilde{x}_1 \mid \tilde{x}_2)_{\tilde{X}} = \tilde{x}_1^* \tilde{x}_2 \quad \text{and} \quad (\tilde{y}_1 \mid \tilde{y}_2)_{\tilde{Y}} = \tilde{y}_1^* \tilde{y}_2
\end{equation*}
and the 2-norms on $ \tilde{X} $ and $ \tilde{Y} $ are also the conventional ones, i.e.
\begin{equation*}
\|\tilde{x}\|_{\tilde{X}}^2 = \|\tilde{x}\|_2^2 \quad \text{and} \quad \|\tilde{y}\|_{\tilde{Y}}^2 = \|\tilde{x}\|_2^2
\end{equation*}

We finally replace the problem matrix $ A $ by the encoding operator $ E $. As stated earlier, its adjoint is given by
\begin{equation*}
E^\dagger = H_X^{-1} E^* H_Y^*
\end{equation*}
Its expression in the tilde coordinates is then given by
\begin{equation*}
\tilde{E} = \psi E \varphi^{-1}
\end{equation*}
and the adjoint in the tilde coordinates is given by
\begin{equation*}
\tilde{E}^\dagger = (\psi E \varphi^{-1})^\dagger = \varphi E^\dagger \psi^{-1} = \varphi \varphi^{-1} (\varphi^*)^{-1} E^* \psi^* \psi \psi^{-1} = (\varphi^*)^{-1} E^* \psi^* = \tilde{E}^*
\end{equation*}
as expected. 

The normal equation for iterative-SENSE can now be written equivalently either in the direct or tilde coordinates as
\begin{equation*}
E^\dagger E x = E^\dagger y_0 \quad \text{or} \quad \tilde{E}^* \tilde{E} \tilde{x} = \tilde{E}^* \tilde{y}_0
\end{equation*}
but this time with arbitrary isomorphisms $ \varphi $ and $ \psi $ for the change of coordinates. If we choose the isomorphisms $ \varphi $ and $ \psi $ as
\begin{equation}
\varphi = I^{-1} \quad \text{and} \quad \psi = \sqrt{D} \tag{$DI2$} \label{eq:DI2}
\end{equation}
it follows
\begin{equation*}
H_X = \varphi^* \varphi = (I^{-1})^* I^{-1} = (I^{-1})^2 \quad \text{and} \quad H_Y = \psi^* \psi = (\sqrt{D})^* \sqrt{D} = D
\end{equation*}
and we recover the original formulation of \cite{iterSense}. However, it is only a special case of the more general formulas in \ref{eq:H3}, and we will from now on consider that $ \varphi $ and $ \psi $ can be any pair of isomorphisms. In the tilde coordinates, $ \varphi $ is the manifestation of ISP and $ \psi $ is the manifestation of DSP. In the direct coordinates, $ H_X $ is the manifestation of ISP and $ H_Y $ is the manifestation of DSP. The same preconditioning manifest itself differently in two different but equivalent worlds. 

We have the following equivalences:
\begin{equation*}
E^\dagger E x^\# = E^\dagger y_0 \quad \Leftrightarrow \quad x^\# \in S_E = \underset{x \in X}{\operatorname{argmin}} \frac{1}{2} \| E x - y_0 \|_{Y,2}^2
\end{equation*}
\begin{equation*}
\tilde{E}^* \tilde{E} \tilde{x}^\# = \tilde{E}^* \tilde{y}_0 \quad \Leftrightarrow \quad \tilde{x}^\# \in S_{\tilde{E}} = \underset{\tilde{x} \in \tilde{X}}{\operatorname{argmin}} \frac{1}{2} \| \tilde{E} \tilde{x} - \tilde{y}_0 \|_2^2
\end{equation*}
and it holds
\begin{equation*}
S_E = \varphi^{-1} S_{\tilde{E}}
\end{equation*}
We note that the standard 2-norm appears in the tilde coordinates, in accordance with our choice in this section.

Given a function 
\begin{equation*}
R(\cdot) \colon X \to \mathbb{R}, \quad x \mapsto R(x)
\end{equation*}
that is well-behaved enough to serve as a regularization function (typically a proper closed convex function), we can now formulate the regularized least-square reconstruction in the direct coordinates as
\begin{equation}
\text{Find } x^\# \in S_E^\text{reg} = \underset{x \in X}{\operatorname{argmin}} \frac{1}{2} \| E x - y \|_{Y,2}^2 + \lambda R(x) \tag{$PER$} \label{eq:PER}
\end{equation}
while in the tilde coordinates it takes the form
\begin{equation}
\text{Find } \tilde{x}^\# \in S_{\tilde{E}}^\text{reg} = \underset{\tilde{x} \in \tilde{X}}{\operatorname{argmin}} \frac{1}{2} \| \tilde{E} \tilde{x} - \tilde{y}_0 \|_2^2 + \lambda \tilde{R}(\tilde{x}) \tag{$P\tilde{E}\tilde{R}$} \label{eq:PtildeEtildeR}
\end{equation}
where
\begin{equation}
\tilde{R}(\tilde{x}) = R(\varphi^{-1} \tilde{x})
\end{equation}
Both problems \ref{eq:PER} and \ref{eq:PtildeEtildeR} are equivalent and their solution sets are related by
\begin{equation*}
S_E^{reg} = \varphi^{-1} S_{\tilde{E}}^{reg}
\end{equation*}
In the discussion section, we discuss the advantage of the direct coordinates over the tilde coordinates.

\section{Experiments}

We present here some experimental tests of our theory on 2D cardiac cinematic (CINE) data with three different kinds of non-cartesian reconstructions described more in detail in the sub-sections below:
\begin{itemize}
    \item a non-regularized reconstruction (iterative-SENSE),  
    \item an $l_1$-spatially regularized reconstruction, 
    \item an $l_1$-temporally regularized reconstruction. 
\end{itemize}
All reconstructions were performed with the Monalisa toolbox \cite{monalisaGit} or with modified functions from that toolbox. It is mainly programmed in Matlab (MathWorks, Natick, Massachusetts, USA) and partially in C++. Please visit the documentation webpage \cite{monalisaDoc} for more information. For all reconstructions, the initial image was obtained by a gridded zero-padded reconstruction (function "bmMathilda" in Monalisa) and we have set the number of iteration to 20 for the iterative reconstructions. In all reconstructions, we have set the data-space preconditioning (DSP) equal to the k-space density compensation (see equations \ref{eq:DI1} and \ref{eq:DI2}), as usually done for non-cartesian reconstructions. Some of our reconstruction were tested with image-space preconditioning (ISP) or without for comparison. Whenever ISP was present, we have set it as given by \ref{eq:DI1} or \ref{eq:DI2} with an additional multiplication by the voxel volume (that multiplication is a scaling due to the implementation choices in Monalisa). Whenever ISP was absent, we have set it equal the voxel volume times the identity. All $l_1$-regularized reconstruction consisted is solving an $l_1$-regularized LS-problem with the alternating direction method of multipliers (ADMM) \cite{admmRef, littleWolf}. Prior to all reconstructions, the coil-sensitivity estimation was performed by the SIGMA method \cite{monalisaDoc, milaniTTV}. 

The goal of our experiments is to validate our theory by demonstrating that
\begin{itemize}
    \item ISP as introduced in \cite{iterSense} can naturally be generalized to least-square regularized reconstructions by implementing it as non-conventional inner-product matrices and in fact accelerates convergence, 
    \item ISP and DSP can equivalently be implemented as non-conventional inner-product matrices or by a linear change of coordinates and both implementations of preconditioning lead to the same result.   
\end{itemize}
After describing the data acquisition and the ground-truth reconstruction in the two coming sub-sections, we present in three next the three kinds of reconstructions used in our experiments, as well as the data preparation for each kind of reconstruction.  

\subsection{Data Acquisition}

 The data used in the present study are available online \cite{dataZenodoo}. These data were all acquired with a 2D single slice gradient echo (GRE) MRI sequence for cardiac CINE imaging on three young and healthy adults on a 3T Prisma scanner (MAGNETOM Prisma, Siemens Healthineers, Erlangen, Germany). Several preliminary images were acquired with localizers in order to place the 2D slice in 4-chamber view on the heart. The acquisition was then performed in end inspiration breath hold with electrocardiogram (ECG) recording for retrospective cardiac gating of the acquired data lines. In order to allow reconstructing a good quality ground-truth, all acceleration strategies were disabled so that a maximum amount of data lines could be acquired while keeping a reasonable breath hold duration for the volunteers. These lead to 256 acquired lines for each frame of the CINE. The acquisition trajectory was radial with 512 points per lines and a (full) field of view (FoV) of 600 mm. We refer the reader to table \ref{table:table_1} for other acquisition parameters. 

In order to estimate the coil-sensitivity by the SIGMA method, some additional calibration scans were performed. They consisted in repeating the main acquisition with the same FoV (in position, size and orientation) once with the the surface coils and once with the body-coil of the scanner. The only difference was that the flip angle was smaller and slice thickness was larger than for the main scan (see table \ref{table:table_1}).  

\begin{table}[ht]
\centering
\begin{tabular}{|l|l|l|}
\hline
\textbf{Parameter} & \textbf{Main Sequence} & \textbf{Calibration Sequence} \\  \textbf{Names} & \textbf{} & \textbf{for Coil-Sensitivities}        \\
\hline
\textnormal{Sequence Type}                     & GRE                                & GRE                                                           \\
\textnormal{Trajectory Type}                   & Radial                             & Radial                                                        \\
\textnormal{Dimensionality}                    & 2D                                 & 2D                                                            \\
\textnormal{TR (Repetition Time)}              & 48.2 \ ms                          & 48.2 \ ms                                                     \\
\textnormal{TE (Echo Time)}                    & 3.8 \ ms                           & 3.8 \ ms                                                      \\
\textnormal{Matrix Size}                       & 512 $\times$ 512 \  (Full \  FOV)  & 512 $\times$ 512 (Full FOV)                                   \\
\textnormal{Flip Angle}                        & 12°                                & \textit{\textbf{5°}}                                          \\
\textnormal{Field of View}                     & 600 $\times$ 600 \  mm² \          & 600 $\times$ 600 \  mm² \                                     \\
\textnormal{Excitation Type}                   & \textnormal{Slice-selective}       & \textnormal{Slice-selective}                                  \\
\textnormal{Pixel Bandwidth}                   & 400 \  Hz/Pix                      & 400 \  Hz/Pix                                                 \\
\textnormal{Number of Lines per Segment}       & 8                                  & 8                                                             \\
\textnormal{Number of Segments}                & 32                                 & 32                                                            \\
\textnormal{Spatial Resolution}                & 1.7 $\times$ 1.7 $\times$ 8.0 \  mm³   & 1.7 $\times$ 1.7 $\times$ \textit{\textbf{20}} \  mm²     \\
\textnormal{Scan Time}                         & 31 s                               & 31 s                                                          \\ 
\hline
\end{tabular}
\caption{\textit{Acquisition parameters for the main sequence (the one leading the data to reconstruct) and the calibration sequence (to perform the coil-sensitivity estimation).}}
\label{table:table_1}
\end{table}

\subsection{Ground-truth Reconstruction}
The totality of 256 data lines acquired for each frame with the radial 2D GRE sequence were used to build the ground-truth images. It means that the raw-data vector $y_0$ consisted of all acquired data-lines. The ground-truth reconstructions were performed with the function "bmSensa" of the Monalisa toolbox, which is an iterative-SENSE implementation solving, the LS-problem 
\begin{equation*}
x^\# \in S_E = \underset{x \in X}{\operatorname{argmin}} \frac{1}{2} \| E x - y_0 \|_{Y,2}^2
\end{equation*}
This reconstruction was performed frame by frame. It can therefore be considered as a static reconstruction performed for each frame individually. In Monalisa, the preconditioning is natively implemented in the inner-product matrices. We chose ISP to be absent for the present ground-truth reconstruction. 

\subsection{The Non-Regularized Reconstruction}
The non-regularized reconstructions were also performed with the function "bmSensa" of Monalisa. Since it is a static reconstruction performed for each frame individually, only one frame was selected and reconstructed in order to compare different variants of the non-regularized reconstruction. Moreover, 50\% of the ground-truth data for that frame (every two data lines) were included for reconstruction. It represents an acceleration factor of 2 w.r.t. ground-truth. The optimization problem solved for that reconstruction was therefore also
\begin{equation*}
x^\# \in S_E = \underset{x \in X}{\operatorname{argmin}} \frac{1}{2} \| E x - y_0 \|_{Y,2}^2
\end{equation*}
but the operator $E$, the data vector $y_0$ and the norm $\| \cdot \|_{Y,2}$ were different as those used in the ground-truth reconstruction because only a part of the trajectory was used and only one frame was reconstructed. The non-regularized reconstruction was tested with and without ISP for comparison. If present, ISP was implemented as \ref{eq:DI1} with an additional multiplication by the voxel volume $\Delta R$ because of the implementation conventions used in Monalisa. 

For comparison, the same reconstruction was modified to implement both ISP and DSP as a linear change of coordinates (as given by \ref{eq:DI2}). The resulting reconstruction function solved the following problem in the tilde-coordinates:  
\begin{equation*}
\tilde{x}^\# \in S_{\tilde{E}} = \underset{\tilde{x} \in \tilde{X}}{\operatorname{argmin}} \frac{1}{2} \| \tilde{E} \tilde{x} - \tilde{y}_0 \|_2^2
\end{equation*}
where 
\begin{equation*}
\tilde{E} \coloneqq \psi E \varphi^{-1}, \quad \tilde{x} \coloneqq \varphi x, \quad \tilde{y}_0 \coloneqq \psi y_0 
\end{equation*}
After that optimization problem was solved, the final image was recovered by
\begin{equation*}
x^\# = \varphi^{-1} \tilde{x}^\#
\end{equation*}

\subsection{The $l_1$-spatially regularized reconstruction}
Since the $l_1$-spatially regularized reconstruction tested in the present article is a static reconstruction that reconstruct each frame individually, a single frame was selected and reconstructed to compare different variant of that reconstruction. The data preparation consisted in randomly selecting 85 of the 256 lines acquired in total for the selected frame, which represent 33\% of the ground-truth data (i.e. acceleration by factor 3 w.r.t. ground-truth). 

The $l_1$-spatially regularized reconstructions were then performed with the function "bmSteva" of Monalisa. We refer the reader to the implementation \cite{monalisaDoc} for a complete information. For short, that reconstruction solves the optimization problem
\begin{equation*}
x^\# \in S_E^{reg} \coloneqq \underset{x \in X}{\operatorname{argmin}} \frac{1}{2} \| E x - y_0 \|_{Y,2}^2 + \frac{\lambda}{2} \| \theta \cdot x \|_{X \times X, 1} 
\end{equation*}
where all quantities in the first term are defined as in the non-regularized reconstruction, and the quantities in the second term (regularization term) are defined as follows. $\lambda$ is a positive regularization parameter and $\| \cdot \|_{X \times X, 1}$ is a 1-norm defined on $X \times X$ as
\begin{equation*}
\| (a, b) \|_{X \times X, 1} \coloneqq \Delta R \cdot  \| a\|_1 + \Delta R \cdot  \| b\|_1
\end{equation*}
where $\| \cdot \|_1$ is the conventional 1-norm for and $a, b \in X$ can be any images. The linear map $\theta$ is here defined to be the finite-difference spatial gradient operator. It sends any image $x$ to the pair of images $(\nabla_{r, 1}x , \nabla_{r, 2}x)$ where $\nabla_{r, 1}x$ is the finite difference derivative in the first spatial dimension and $\nabla_{r, 2}x$ is the one in the second spatial dimension. Since $\nabla_{r, 1}x$ and $\nabla_{r, 2}x$ have the same size like an image, it is natural to consider them as images as well. Formally, the map $\theta$ is thus given by
\begin{align*}
\theta : \  X & \longrightarrow X \times X \\
x & \longmapsto \left( \nabla_{r, 1}x, \nabla_{r, 2}x \right) \in X \times X 
\end{align*}
This $l_1$-regularized reconstruction was tested with and without ISP for comparison, with ISP implemented as inner-product matrices if present (as given by \ref{eq:DI1}).  

For comparison with the other implementation strategy of preconditioning, namely by a linear change of coordinates, the bmSteva function was modified. The changes of coordinate were set as given by \ref{eq:DI2}. The resulting reconstruction function solved then the optimization problem
\begin{equation*}
\tilde{x}^\# \in S_{\tilde{E}}^{reg} \coloneqq \underset{\tilde{x} \in \tilde{X}}{\operatorname{argmin}} \frac{1}{2} \| \tilde{E} \tilde{x} - \tilde{y}_0 \|_2^2 + \frac{\lambda}{2} \| \tilde{\theta} \cdot \tilde{x} \|_{\tilde{X} \times \tilde{X}, 1} 
\end{equation*}
where 
\begin{equation*}
\tilde{E} \coloneqq \psi E \varphi^{-1}, \quad \tilde{x} \coloneqq \varphi x, \quad \tilde{y}_0 \coloneqq \psi y_0 
\end{equation*}
In order to define the map $\tilde{\theta}$, we first need to define the map $\left( \varphi, \varphi \right)$ as follows 
\begin{align*}
\left( \varphi, \varphi \right) : \  X \times X & \longrightarrow \tilde{X} \times \tilde{X} \\
(a, b) & \longmapsto \left( \tilde{a}, \tilde{b} \right) \coloneqq \left( \varphi a , \varphi b \right)
\end{align*}
or in other words
\begin{align*}
\left( \varphi, \varphi \right)(a, b) = \left( \varphi a , \varphi b \right) = \left( \tilde{a}, \tilde{b} \right) 
\end{align*}
We define then the map $\tilde{\theta}$ by 
\begin{equation*}
\tilde{\theta} \coloneqq \left( \varphi, \varphi \right) \circ   \theta \circ  \varphi^{-1} 
\end{equation*}
which is to that the following diagram commutes: 
\begin{equation*}
\begin{tikzcd}
X \arrow{r}{\theta} \arrow[swap]{d}{\varphi}  &  X \times X \arrow{d}{\left( \varphi, \varphi \right)} \\
\tilde{X} \arrow{r}{\tilde{\theta}}   &    \tilde{X} \times \tilde{X} 
\end{tikzcd}
\end{equation*}
In order to ensure that the regularization in the tilde-coordinates is the same as in the direct coordinates, the 1-norm on $\tilde{X} \times \tilde{X}$ has to be defined as
\begin{equation*}
\| ( \tilde{a}, \tilde{b} ) \|_{\tilde{X} \times \tilde{X}, 1} \coloneq \Delta R \cdot  \| \varphi^{-1} \tilde{a}\|_1 + \Delta R \cdot  \| \varphi^{-1} \tilde{b} \|_1
\end{equation*}
where $\| \cdot \|_1$ is the conventional 1-norm. After the reconstruction is performed in the tilde-coordinates (by solving the previous optimization problem), the final image can then be recovered by
\begin{equation*}
x^\# = \varphi^{-1} \tilde{x}^\#
\end{equation*}

\subsection{The $l_1$-temporally regularized reconstruction}

The $l_1$-temporally regularized reconstruction is a multiple frame reconstruction that shares information between frames to fill potentially missing data. It is why the data of all frames were included to test that reconstruction. The data preparation consisted in randomly selecting for each frame 40 of the 256 lines acquired, which represent 15.6\% of the ground-truth data (i.e. acceleration by factor 6.4 w.r.t. ground-truth).

We will call $nVPF$ the number of voxels per frame (i.e. the number of voxels in one frame) and $nFr$ the number of frames, so that the total number of voxels $nVox$ is equal to $nVPF \cdot nFr$. In the present case, the image space is then  
\begin{equation*}
X = X^{(1)} \times ... \times X^{(nFr)} \simeq \mathbb{C}^{nVPF \cdot nFr} =  \mathbb{C}^{nVox}
\end{equation*}
where each frame-space $X^{(i)}$ verifies $X^{(i)} \simeq \mathbb{C}^{nVPF}$ for $i \in {1, ..., nFr}$. An image $x \in X$ is then defined as a vertical catenation 
\begin{equation*}
x = \begin{bmatrix} x^{(1)} \\  \vdots \\ x^{(nFr)} \end{bmatrix} \in X
\end{equation*}
where each frame $x^{(i)}$ is an element of $X^{(i)}$ for $i \in {1, ..., nFr}$.

Similarly, any vector $y$ is the vertical catenation of $nFr$ data-bins $y^{(1)}, ..., y^{(nFr)}$ where each data-bin $y^{(i)}$ is a column vector of $nPt^{(i)} \cdot nCh $ complex numbers ($nPt$ stands for "number of points" and $nCh$ the "number of channels"). We will write $Y^{(i)} \simeq \mathbb{C}^{nPt^{(i)}\cdot nCh}$ for the data-bin space number $i$ so that we can write $y^{(i)} \in Y^{(i)}$. 
The data-space is then given by $Y = Y^{(1)} \times ... \times Y^{(nFr)}$. The data vector $y_0$ is then vertical catenation of the $nFr$ vectors $y_0^{(1)} \in Y^{(1)}$, ..., $y_0^{(nFr)} \in Y^{(nFr)}$:  
\begin{equation*}
y_0 = \begin{bmatrix} y_0^{(1)} \\  \vdots \\ y_0^{(nFr)} \end{bmatrix} \in Y
\end{equation*}

The $l_1$-temporally regularized reconstructions were then performed with the function "bmTevaDuoMorphosia\_chain" of Monalisa. We refer the reader to the implementation \cite{monalisaDoc} for a complete information. For short, that reconstruction solves the optimization problem
\begin{equation*}
x^\# \in S_E^{reg} \coloneqq \underset{x \in X}{\operatorname{argmin}} \frac{1}{2} \| E x - y_0 \|_{Y,2}^2 + \frac{\lambda}{2} \| \theta \cdot x \|_{X \times X, 1} 
\end{equation*}
Here is the encoding operator $E$ defined by the digonal block matrix
\begin{equation*}
E = \underbrace{\begin{bmatrix}     F^{(1)}C    &        &   
                    \\              & \ddots &    
                    \\              &        & F^{(nFr)}C \end{bmatrix}}_{\substack{nFr \  Blocks}}
\end{equation*}
 where $F^{(i)}$ is the discrete non-uniform Fourier transform associated with the trajectory of data bin number $i$. Further is the 2-norm $\| \cdot \|_{Y, 2}$ defined by
 \begin{equation*}
\| y \|_{Y, 2} \coloneq  y^* H_Y y
 \end{equation*}
where the inner product matrix $H_Y$ is here defined by
\begin{equation*}
H_Y = \underbrace{\begin{bmatrix}     D^{(1)}    &        &   
                    \\               & \ddots &    
                    \\               &        & D^{(nFr)} \end{bmatrix}}_{\substack{nFr \  Blocks}}
\end{equation*}
with $D^{(i)}$ being the k-space density compensation of the trajectory of data-bin number $i$. This definition of $H_Y$ is the embedding of DSP in the inner-product matrix. 

The 2-norm on the image-space $X$ is given similarly by
 \begin{equation*}
\| x \|_{X, 2} \coloneq  x^* H_X x
 \end{equation*}
This $l_1$-regularized reconstruction was tested with and without ISP for comparison. To test it in absence of ISP, the inner product matrix $H_X$ was defined as
\begin{equation*}
H_X \coloneq \Delta R \underbrace{\begin{bmatrix}     id_{X^{(1)}}     &        &   
                    \\                                          & \ddots &    
                    \\                                          &        & id_{X^{(nFr)}} \end{bmatrix}}_{\substack{nFr \  Blocks}}
\end{equation*}
where $id_{X^{(i)}}$ is the identity map on $X^{(i)}$. To test it with the presence of ISP, it was defined as
\begin{equation*}
H_X \coloneq \Delta R \underbrace{\begin{bmatrix}     \left(I^{-1}\right)^2     &        &   
                    \\                              & \ddots        &    
                    \\                              &               & \left(I^{-1}\right)^2 \end{bmatrix}}_{\substack{nFr \  Blocks}}
\end{equation*}

The embedding of ISP and DSP in the inner-product matrices presented here (as implemented in the native version of "bmTevaDuoMorphosia\_chain") corresponds to the choice expressed in \ref{eq:DI1}, up to a multiplication by $\Delta R$ that is necessary to fit the conventions of the Monalisa toolbox. 
 
 $\lambda$ is a positive regularization parameter and $\| \cdot \|_{X \times X, 1}$ is a 1-normed defined on $X \times X$ as
\begin{equation*}
\| (a, b) \|_{X \times X, 1} \coloneqq \Delta R \cdot  \| a\|_1 + \Delta R \cdot  \| b\|_1
\end{equation*}
where $\| \cdot \|_1$ is the conventional 1-norm for and $a, b \in X$ can be any images. The linear map $\theta$ is here defined to be the pair of the forward and backward finite-difference time derivative: 
\begin{align*}
\theta : \  X & \longrightarrow X \times X \\
x & \longmapsto \left( D_t^b x, D_t^f x \right) \in X \times X 
\end{align*}
Since the finite-difference time derivative of an image has the dimension of an image, it is natural to consider it as an image as well. We define $D_t^b$ and $D_t^f$ by the block-matrices

\begin{equation*}
D_t^f = \Delta R    \begin{bmatrix}     id_{X^{(1)}}  & -T^{(1, 2)}  & 0      & \cdots & 0   
                    \\                  0             & \ddots       & \ddots & \ddots & \vdots
                    \\                  \vdots        & \ddots       & \ddots & \ddots & 0
                    \\                  0             & \ddots       & \ddots & \ddots & -T^{(nFr-1, nFr)}
                    \\                  -T^{(nFr, 1)} & 0            & \cdots & 0      & id_{X^{(nFr)}} \end{bmatrix}
\end{equation*}
and
\begin{equation*}
D_t^b = \Delta R    \begin{bmatrix}     id_{X^{(1)}}  & 0      & \cdots & 0 & -T^{(nFr, 1)}   
                    \\                  -T^{(2, 1)}   & \ddots & \ddots & \ddots & 0
                    \\                  0             & \ddots & \ddots & \ddots & \vdots
                    \\                  \vdots        & \ddots & \ddots & \ddots & 0
                    \\                  0             & \cdots & 0 & -T^{(nFr, nFr-1)}      & id_{X^{(nFr)}} \end{bmatrix}
\end{equation*}
Here is matrix $T^{(i, j)}$ the deformation matrix which is so that $T^{(i, j)}\cdot x^{(j)}$ matches (is superimposed to) $x^{(i)}$, up to some noise, registration error and other errors. We can see $T^{(i, j)}$ as a linear map of the form
\begin{align*}
& T^{(i, j)} : X^{(j)} \longrightarrow X^{(i)} \\
& x^{(j)} \longmapsto T^{(i, j)} \cdot x^{(j)} 
\end{align*}
Since the deformation matrices are unknown before the reconstruction has begun, a first reconstruction with 20 iterations is performed with bmTevaDuoMorphosia\_chain by setting $T^{(i, j)}$ equal to the map $id^{(i, j)}$ that we define to transport $x^{(j)}$ from $X^{(j)}$ to $X^{(i)}$ without changing any of its entries. We will write that map
\begin{align*}
& id^{(i, j)} : X^{(j)} \longrightarrow X^{(i)} \\
& x^{(j)} \longmapsto x^{(j)} 
\end{align*}
Then, an estimation of the deformation fields between each $x^{(i)}$ and its forward and backward temporal neighboors was performed with the "imregdeamon" registration tool of MATLAB. The deformation fields were encoded in the (sparse) matrices $T^{(i+1, i)}$ and $T^{(i-1, i)}$ and the reconstruction was run again with 20 iterations with these new deformation matrices.  

This $l_1$-regularized reconstruction was tested with and without ISP for comparison, with ISP implemented in the inner-product matrices as described above.  

For comparison with the other implementation strategy of preconditioning, namely by a linear change of coordinates, the "bmTevaDuoMorphosia\_chain" function was modified. The changes of coordinate were set by 
\begin{equation*}
\varphi \coloneq \sqrt{H_X} \quad\quad \psi \coloneq \sqrt{H_Y}  
\end{equation*}
where the square root of matrices is to be interpreted as a componentwise square root (which makes sense since $H_X$ and $H_Y$ are both diagonal with positive entries). This definition is similar to \ref{eq:DI2} but takes account for multiple frames. The resulting reconstruction function solved then the optimization problem in the tilde coordinate: 
\begin{equation*}
\tilde{x}^\# \in S_{\tilde{E}}^{reg} \coloneqq \underset{\tilde{x} \in \tilde{X}}{\operatorname{argmin}} \frac{1}{2} \| \tilde{E} \tilde{x} - \tilde{y}_0 \|_2^2 + \frac{\lambda}{2} \| \tilde{\theta} \cdot \tilde{x} \|_{\tilde{X} \times \tilde{X}, 1} 
\end{equation*}
where 
\begin{equation*}
\tilde{E} \coloneqq \psi E \varphi^{-1}, \quad \tilde{x} \coloneqq \varphi x, \quad \tilde{y}_0 \coloneqq \psi y_0 
\end{equation*}
In order to define the map $\tilde{\theta}$, we proceed as for the $l_1$-spatially regularized reconstruction. We first define the map $\left( \varphi, \varphi \right)$ by 
\begin{align*}
\left( \varphi, \varphi \right) : \  X \times X & \longrightarrow \tilde{X} \times \tilde{X} \\
(a, b) & \longmapsto \left( \tilde{a}, \tilde{b} \right) \coloneqq \left( \varphi a , \varphi b \right)
\end{align*}
or in other words
\begin{align*}
\left( \varphi, \varphi \right)(a, b) = \left( \varphi a , \varphi b \right) = \left( \tilde{a}, \tilde{b} \right) 
\end{align*}
Then we define then the map $\tilde{\theta}$ by 
\begin{equation*}
\tilde{\theta} \coloneqq \left( \varphi, \varphi \right) \circ   \theta \circ  \varphi^{-1} 
\end{equation*}
which is to that the following diagram commutes: 
\begin{equation*}
\begin{tikzcd}
X \arrow{r}{\theta} \arrow[swap]{d}{\varphi}  &  X \times X \arrow{d}{\left( \varphi, \varphi \right)} \\
\tilde{X} \arrow{r}{\tilde{\theta}}   &    \tilde{X} \times \tilde{X} 
\end{tikzcd}
\end{equation*}
In order to ensure that the regularization in the tilde-coordinates is the same as in the direct coordinates, the 1-norm on $\tilde{X} \times \tilde{X}$ has to be defined as
\begin{equation*}
\| ( \tilde{a}, \tilde{b} ) \|_{\tilde{X} \times \tilde{X}, 1} \coloneq \Delta R \cdot  \| \varphi^{-1} \tilde{a}\|_1 + \Delta R \cdot  \| \varphi^{-1} \tilde{b} \|_1
\end{equation*}
where $\| \cdot \|_1$ is the conventional 1-norm. After the reconstruction is performed in the tilde-coordinates (by solving the previous optimization problem), the final image can then be recovered by
\begin{equation*}
x^\# = \varphi^{-1} \tilde{x}^\#
\end{equation*}

\section{Results}

\begin{figure} 
    \centering
    \includegraphics[width=\linewidth]{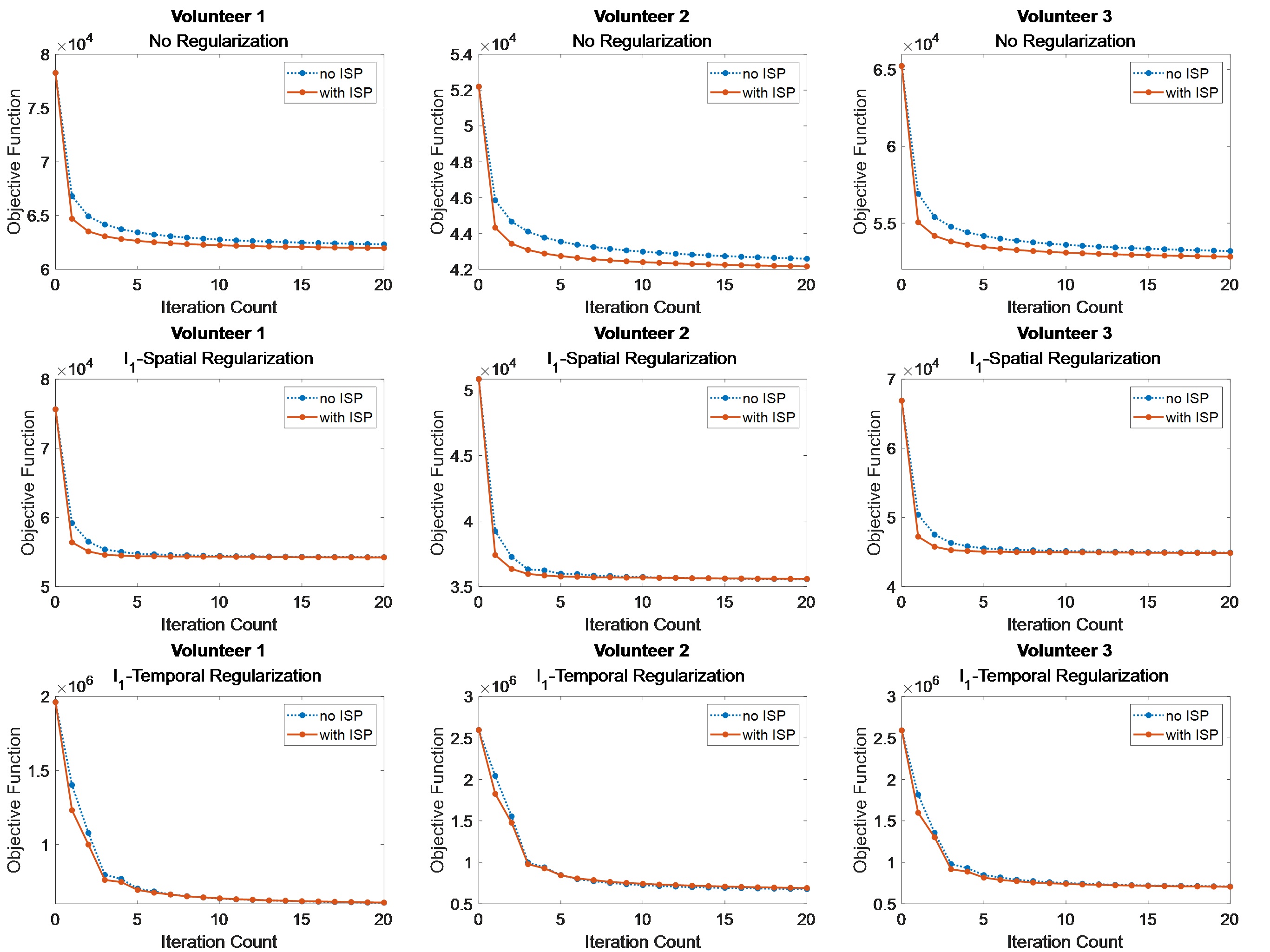}
    \caption{\textit{This figure displays the values of the objective function for each iteration and each volunteer, and for each of the three chosen types of reconstruction. The orange solid line is for the reconstruction including image space preconditioning (ISP), and the blue pointed line is for the reconstruciton without ISP. There is one column for each volunteer. The first row of subfigures is for the non-regularized reconstruction, the second for the $l_1$-spatially regularized reconstruction, and the third for the $l_1$-temporally regularized reconstruction}}
    \label{fig:figure_objVal}
\end{figure}

The values of the objective function are displayed for each reconstruction and each iteration in figure \ref{fig:figure_objVal}. The first column is for volunteer 1, the second for volunteer 2 and the third for volunteer 3. The blue line corresponds to reconstructions without image-space preconditioning (ISP), and the orange line corresponds to reconstructions with ISP. The first row of sub-figures shows that for the reconstruction without regularization (i.e. iterative-SENSE), preconditioning improves convergence speed, as expected. The same effect is observed for the reconstruction with $l_1$-spatial regularization (middle row of sub-figures) even if the improvement brought by ISP is smaller than for non-regularized reconstructions. The third row of sub figure show that ISP only brings a very small improvement in the convergence speed for $l_1$-temporally regularized reconstructions. Moreover, the value of the objective function for volunteer 2 converges two a slightly larger value when using ISP as compared to an absence of ISP. However, for all three volunteers, ISP does improve the convergence speed for $l_1$-temporally regularized reconstructions despite its very small effect. Even for volunteer 2, the values of the objective function reaches the plateau faster with ISP than in absence of ISP, although the objective function values are slightly more favorable in absence of ISP.        

\begin{figure}
    \centering
    \includegraphics[width=\linewidth]{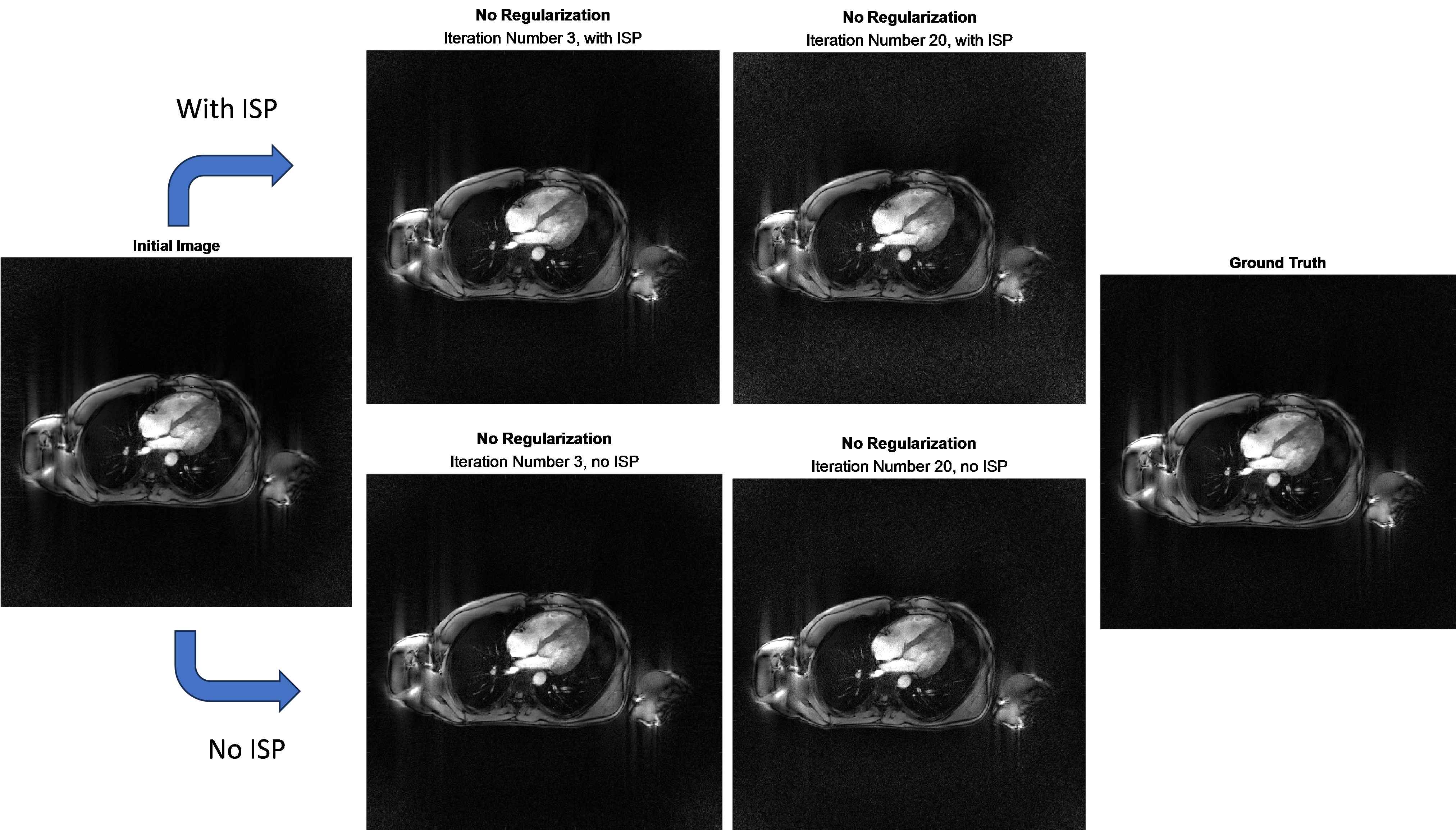}
    \caption{\textit{This figure displays the reconstructed image at iteration number 3 and 20 (last iteration) for the non-regularized reconstruction with use of image space preconditioning (ISP, upper row of subfigures) and without use of ISP (lower row of subfigures). The initial image for both reconstructions is displayed on the left and the ground truth on the right. This reconstruction include 50 \% of the data used to generate the ground-truth. We notice that while ISP increases convergence speed, it also accelerate the emerging of noise along iterations, a known phenomenon for non-regularized reconstructions. }}
    \label{fig:figure_reconNonReg}
\end{figure}

Figure \ref{fig:figure_reconNonReg} shows the initial image, the image a iteration number 3, and the image at iteration number 20 for the non-regularized reconstruction (iterative-SENSE) for which we chose to include every two data lines of the ground-truth reconstruction (under-sampling factor 2 with respect to ground-truth). The upper row of sub-figures corresponds the reconstruction with ISP, and the lower row corresponds to the reconstruction without ISP. The ground-truth is displayed on the right of the figure. We notice that for both reconstructions (with and without ISP), noise is amplified along iterations, and that this effect is stronger when using ISP. 

\begin{figure}
    \centering
    \includegraphics[width=\linewidth]{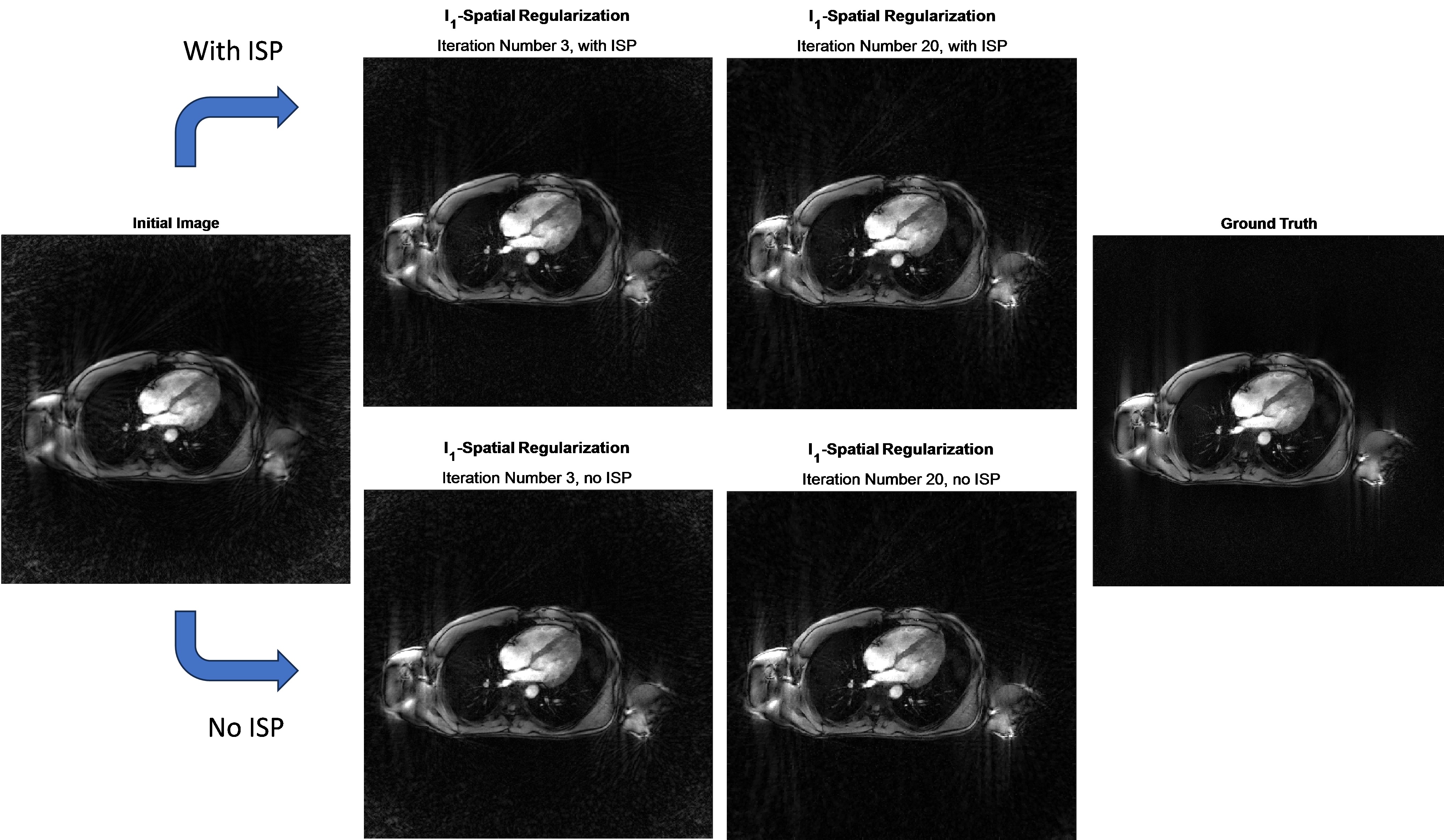}
    \caption{\textit{This figure displays the reconstructed image at iteration number 3 and 20 (last iteration) for the $l_1$-spatially regularized reconstruction with use of image space preconditioning (ISP, upper row of subfigures) and without use of ISP (lower row of subfigures). The initial image for both reconstructions is displayed on the left and the ground truth on the right. This reconstruction includes 33.2 \% of the data used to generate the ground-truth. In contrast to the non-regularized reconstruction, this  $l_1$-spatially regularized reconstruction do not suffer of the emerging of noise along iterations. Moreover, we observe that undersampling artifacts present in the corners of the image are slightly more reduced at iteration number 3 when ISP is used as compared to when ISP is absent. }}
    \label{fig:figure_reconSpaReg}
\end{figure}

Figure \ref{fig:figure_reconSpaReg} shows the initial image, the image at iteration number 3, and the image at iteration number 20 for the $l_1$-spatially regularized reconstruction for which we chose to randomly include one third (33\%) of the lines of the ground-truth reconstruction (under-sampling factor 3 with respect to ground-truth). The upper row of sub-figures corresponds the reconstruction with ISP, and the lower row corresponds to the reconstruction without ISP. The ground-truth is displayed on the right of the figure. We notice that in both reconstructions (with and without ISP), noise is not amplified as in the non-regularized reconstruction. Moreover, we note that the reconstruction at iteration number 3 is slightly better for the reconstruction with ISP as indicated by fewer under-sampling artifacts in the corners, in accordance with the faster convergence using ISP as demonstrated in figure \ref{fig:figure_objVal} (middle row). 

\begin{figure}
    \centering
    \includegraphics[width=\linewidth]{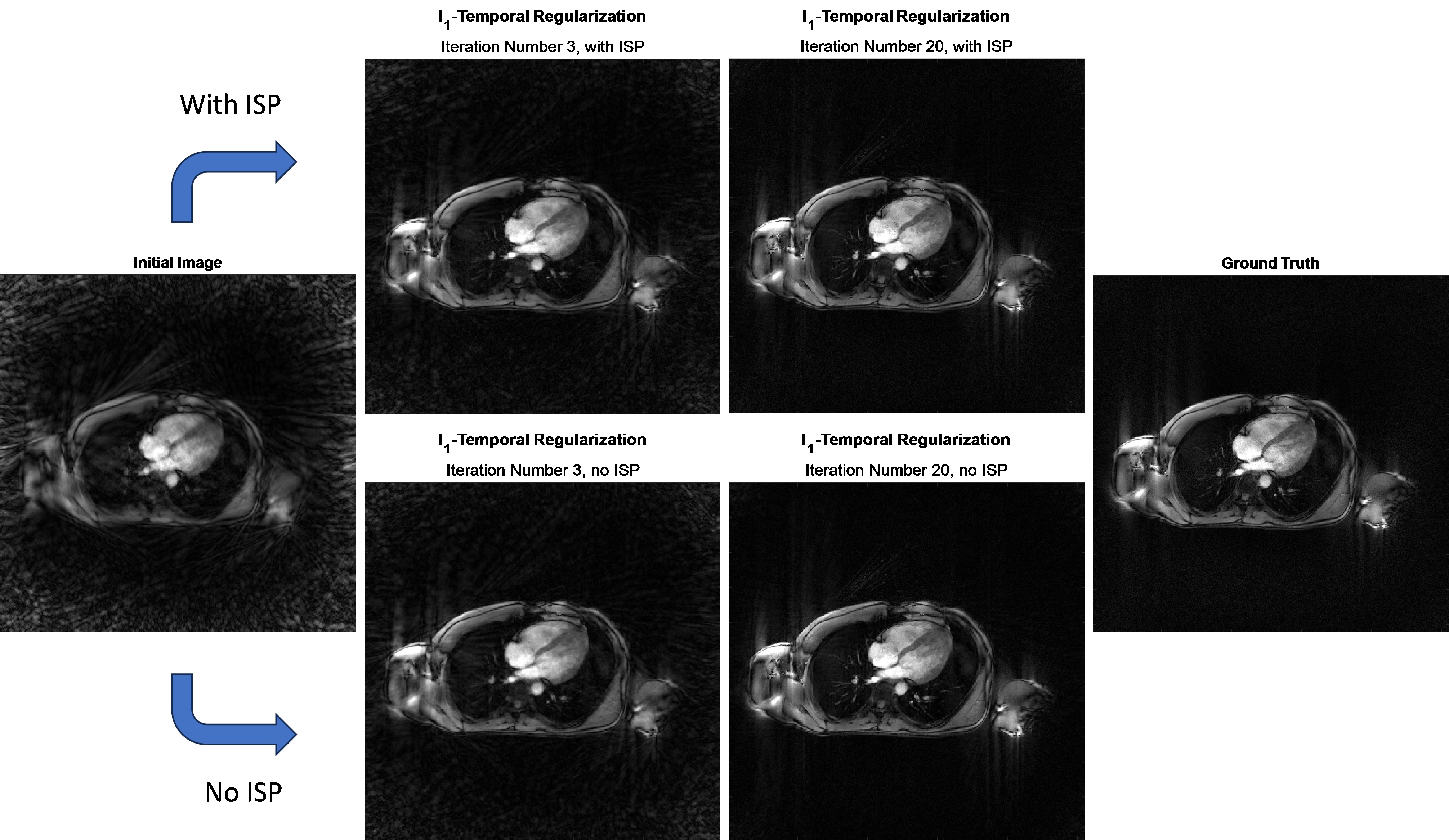}
    \caption{\textit{This figure displays the reconstructed image at iteration number 3 and 20 (last iteration) for the $l_1$-temporally regularized reconstruction with use of image space preconditioning (ISP, upper row of subfigures) and without use of ISP (lower row of subfigures). This reconstruction included only 15.5 \% of the data used for to generate the ground-truth. The initial image for both reconstructions is displayed on the left and the ground truth on the right. We notice here no visible difference on the reconstructed images between the presence of absence of ISP. One can however appreciate the improvement of image quality from the initial image to the last iteration (number 20) }}
    \label{fig:figure_temReg}
\end{figure}

Figure \ref{fig:figure_temReg} shows the initial image, the image at iteration number 3, and the image at iteration number 20 for the $l_1$-temporally regularized reconstruction for which we chose to randomly include 15.5 \% of the lines of the ground-truth reconstruction (under-sampling factor 6.5 with respect to ground-truth). The upper row of sub-figures corresponds the reconstruction with ISP, and the lower row corresponds to the reconstruction without ISP. The ground-truth is displayed on the right of the figure. We notice that both reconstructions (with and without ISP) shows a significant improvement along iterations and the last image (iteration 20) is visually close to the ground truth. However, no acceleration of convergence due to ISP is visually visible on those figure, in accordance with the values of the objective function in figure \ref{fig:figure_objVal} (lower row). 

Figures \ref{fig:figure_objVal}, \ref{fig:figure_reconNonReg}, \ref{fig:figure_reconSpaReg} and \ref{fig:figure_temReg} were all obtained by implementing preconditioning in the inner-product matrices. All reconstructions were also successfully implemented by integrating  ISP and DSP a linear change of coordinates. 

\begin{figure}
    \centering
    \includegraphics[width=\linewidth]{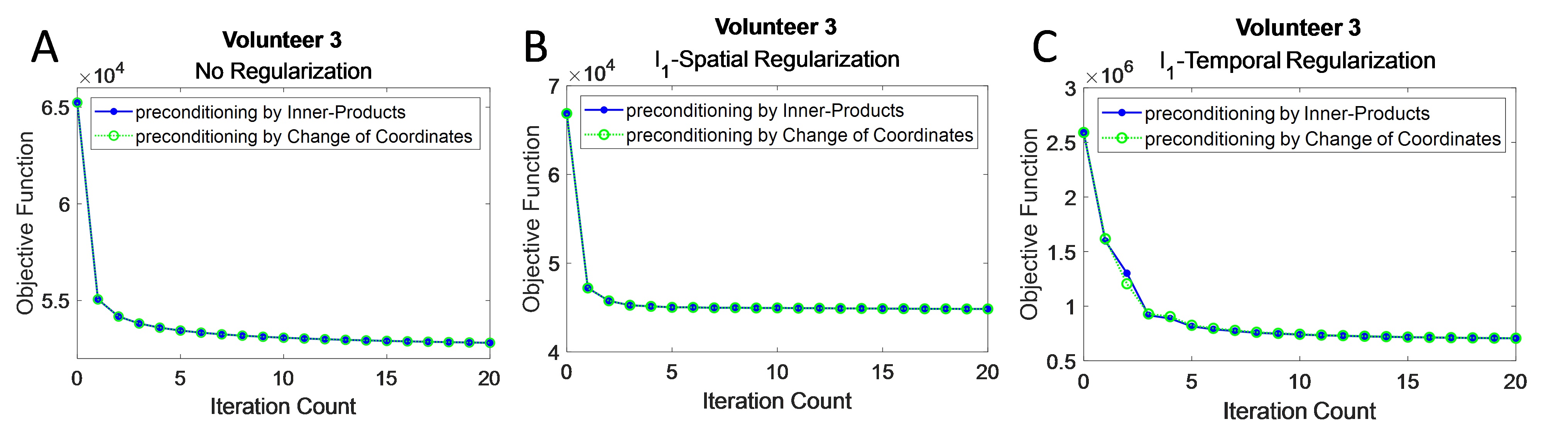}
    \caption{\textit{This figure displays the values of the objective function at each iteration for the three kind of tested reconstructions, on volunteer 3, and with use of image-space preconditioning (ISP). The blue solid line stands for the reconstructions with ISP implemented in the inner-product matrices, and the green dashed lines stands for the reconstructions with ISP implemented by a linear change of coordinates. We see that for the non-regularized reconstruction and the $l_1$-spatially regularized reconstruction, there are no visible differences in those objective function values between both types of implementation. For the $l_1$-temporally regularized reconstruction however, we do notice a difference in the first iteration, which disappear in the next iterations. }}
    \label{fig:figure_psiPhi}
\end{figure}

For the reconstruction without regularization, the exact same images where obtained with preconditioning by change of coordinates, up to an average difference (over voxels) of the order of $10^{-7}$, which is a machine epsilon for single precision. Figure \ref{fig:figure_psiPhi} A shows that the values of the objective function for both implementations of preconditioning are super-imposed, for the reconstruction without regularization. 

For the reconstruction with $l_1$-spatial regularization, the images obtained with preconditioning by change of coordinates were very close to the images obtained with preconditioning by inner-product matrices: the average difference (over voxels) was of the order of $10^{-4}$, which is we can still attribute to a difference in the truncations errors. Figure \ref{fig:figure_psiPhi} B shows that the values of the objective function for both implementation of preconditioning are super-imposed, for the reconstruction with $l_1$-spatial regularization. 

\begin{figure}
    \centering
    \includegraphics[width=\linewidth]{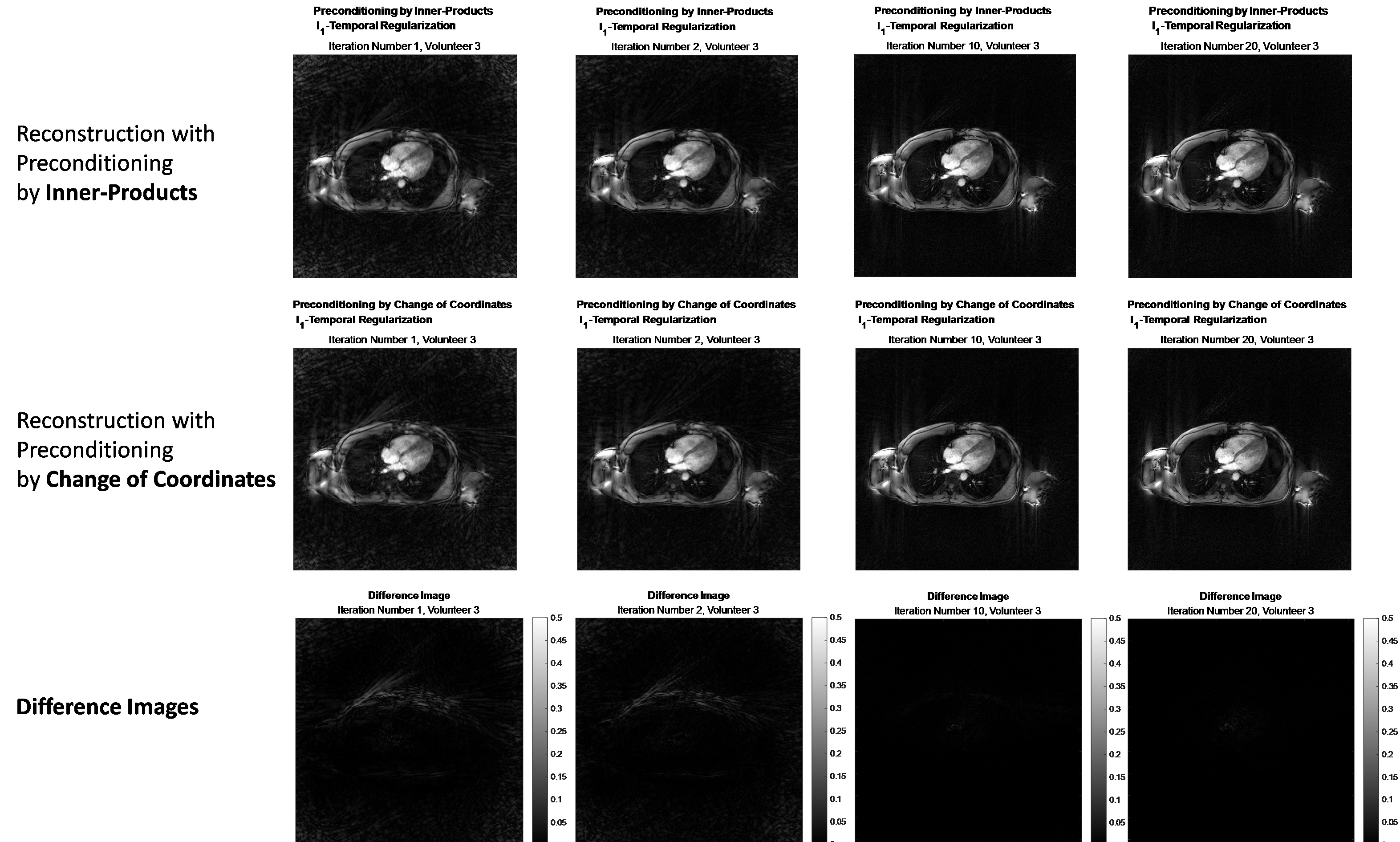}
    \caption{\textit{This figure displays the reconstructed image at iteration 2, 3, 10 and 20 for the $l_1$-temporally regularized reconstruction with image-space preconditioning (ISP) implemented in the inner-product matrices (upper row of subfigures) and with ISP implemented by a linear change of coordinates (lower row of subfigures). The third row of subfigures is simply the difference image between the first and second row. We do notice that both implementation of ISP result in slightly different reconstructed images in the few first iterations. }}
    \label{fig:figure_diffIm}
\end{figure}

For the reconstruction with $l_1$-temporal regularization, the images obtained with preconditioning by change of coordinates were slightly different from the images obtained with preconditioning by inner-product matrices for the first iterations, as shown for volunteer 3 in figure \ref{fig:figure_diffIm}. However, the final images were undistinguishable by the human eye since the average difference (over voxels) between both implementations of preconditioning was of the order of 0.0025 (the gray value of blood being around 1). Figure \ref{fig:figure_psiPhi}C shows that the values of the objective function for both implementation of preconditioning are practically super-imposed after a few iterations for the reconstruction with $l_1$-temporal regularization 

\section{Discussion}

In this article, we showed that ISP and DSP could both be introduced directly in the variational formulation of the reconstruction problem (i.e. in an algorithm-independent manner) in two different but equivalent manners: either by problem \ref{eq:PtildeEtildeR} 
\begin{equation*}
\text{Find } x^\# \in \varphi^{-1} \cdot \underset{\tilde{x} \in \tilde{X}}{\operatorname{argmin}} \frac{1}{2} \| \tilde{E} \tilde{x} - \tilde{y}_0 \|_2^2 + \lambda \tilde{R}(\tilde{x}) 
\end{equation*}
with conventional inner-products, or alternatively by problem \ref{eq:PER}
\begin{equation*}
\text{Find } x^\# \in  \underset{x \in X}{\operatorname{argmin}} \frac{1}{2} \| E x - y \|_{Y,2}^2 + \lambda R(x) 
\end{equation*}
with non-conventional inner-products $(\cdot\vert\cdot)_X$ resp. $(\cdot\vert\cdot)_Y$ defined on $X$ resp. $Y$. Any algorithm solving those reconstruction problem should in principle be preconditioned by ISP and DSP without needing to be inject preconditioning in the algorithms subsequently. This provides a systematic method to insert ISP and to understand how ISP has to be implemented in algorithm, which is not obvious if ISP is injected in the algorithm artificially. Among the two manners, the second is more natural because it allows to work with the reconstructed image $x$ itself instead of a substitute variable and it does not need a mulatiplication by $\varphi^{-1}$ after reconstruction to recover the final image. 

Although the second formulation \ref{eq:PER} is more natural, the state of the art in MRI reconstruction is to formulate the reconstruction problem as a simplification of the first formulation \ref{eq:PtildeEtildeR}. In that setting, conventional inner-products are used and a (partial) change of coordinates is performed:  $\psi$ is set equal to the square root of a matrix $\sqrt{D}$ that is often set equal to the k-space density compensation or some other matrix such as the GROG weighting-function \cite{graspPro, graspGrog} (which are two special cases of DSP), and $\varphi$ is set equal to the identity. It results that $x$ equals $\tilde{x}$, $x^{\#}$ equals $\tilde{x}^{\#}$, $\tilde{R}(\tilde{x})$ equals $R(x)$, and $X$ equals $\tilde{X}$ (equals $\mathbb{C}^{nVox}$). This is why all recent formulations of the MRI reconstruction problem are stated in a declination of the form:
\begin{equation*}
\text{Find }x^{\#} \in \underset{x \in X}{\operatorname{argmin}} \frac{1}{2} \left\| \sqrt{D} E x - \sqrt{D} y_0 \right\|_2^2  + \lambda R(x)
\end{equation*}
with conventional inner-products. This way of formulating the reconstruction problem does not include ISP, and as a matter of fact, many published methodologies usually do not perform ISP in the reconstruction algorithms \cite{noPrecond1, noPrecond2, noPrecond3, noPrecond4}. 

The reason why the less natural formulation imposed itself in the MRI community and why ISP was not introduced in the reconstruction problem is somewhat mysterious because ISP was originally present at the source \cite{iterSense}. A possibility, although hypothetical, is the following: non-conventional inner-products are probably a novelty in the field of MRI reconstruction and since researchers had to formulate the reconstruction problem without it, they chose problem \ref{eq:PtildeEtildeR} because it uses conventional inner-products. In addition, since working with a substitute variable instead of the image is not natural and needs a final multiplication by $\varphi^{-1}$ after reconstruction, they discarded ISP by setting $\varphi$ equal to the identity. It resulted the disappearing of ISP from the variational problems, which reappeared in some algorithms \cite{ong2020, noPrecond1, noPrecond2, noPrecond3, noPrecond4} but did not propagate to all iterative algorithms. The present text is an attempt to solve that issue. This can potentially impact all modern iterative reconstructions, including iterative deep-learning reconstructions. 

We showed in some numerical experiments that what was claimed in the theory section indeed works in practice. The convergence speed is obviously increased by ISP for the chosen examples of non-regularized and $l_1$-spatially regularized reconstructions. Although the $l_1$-temporally reconstruction benefited much less from ISP, the results still show a slightly faster convergence speed, so that the proof of concept holds. Overall, figure \ref{fig:figure_objVal} confirms that the use of ISP increases the convergence speed for three examples of reconstructions tested. 

The fact that the ISP we chose increases convergence speed for the non-regularized reconstruction, but has a very small effect for the temporally regularized reconstruction (which is the most effective of the three reconstructions presented here in order to compensate for undersampling) is not explained and has not been observed in the past, as far as the authors could be informed. We chose here the ISP proposed by Pruessmann in \cite{iterSense} because it was the first one and showing 25 years later that this ISP could be re-introduced in the modern frame work of reconstruction kind of closed a loop. But it would now be interesting to test other kind of ISP as proposed in the literature to see if it can accelerate the convergence of temporally regularized reconstructions. 

\section{Conclusion}

We have introduced two algorithm-independent formulations of data-space preconditioning and image-space preconditioning for MRI reconstruction by introducing them directly in the (abstract) variational formulation of the reconstruction problem. Any algorithm solving the reconstruction problem in one of those formulations should in principle benefits from image-space and data-space preconditioning. It does not need to be integrated artificially into the algorithm implementation. This also holds for heuristic modifications of classical reconstruction algorithms, such as many iterative deep learning reconstructions. This reformulation of preconditioning allows for the natural embedding of image space preconditioning in all modern iterative reconstructions, which has been lacking until now.

We have argued that among the two formulations, one is more natural because it allows a systematic implementation of preconditioning in all algorithms and allows working directly with the reconstructed image instead of a substitute variable. This more natural formulation is however currently absent from the state of the art in MRI reconstruction. We aim to resolve that issue with the present article. Also, by introducing non-conventional inner-product on image-space and data-space, we have restored a symmetry between them which has bin lacking until now. 

Finally, we successfully demonstrated by practical experiments on three volunteers that the implementation of one example of image-space preconditioning following our theory indeed increases convergence speed and that the two formulations of preconditioning (by non-conventional inner-products or by a change of coordinates) indeed lead to the same result in practice.   

\bibliographystyle{IEEEtran}
\bibliography{myBib}

\end{document}